\documentclass{amsart}

\usepackage{amssymb}
\usepackage{enumerate, xspace}
\usepackage{dsfont}
\usepackage{changebar}
\usepackage[dvips]{graphicx}
\usepackage{amsthm}

\newif\ifjoel
\ifjoel\fi

\numberwithin{equation}{section}
\newtheorem{thm}{Theorem}[section]
\newtheorem{lem}[thm]{Lemma}

\newenvironment{proofof}[1]{\medskip\noindent
   \textbf{Proof of #1:} }{\hfill \qed\par\medskip}

\theoremstyle{definition}
\newtheorem{rem}[thm]{Remark}

\newcommand\cL{{\mathcal L}}

\newcommand\cZ{{\mathcal Z}}

\newcommand\bZ{{\mathbb Z}}

\newcommand{\be}{\begin{equation}}
\newcommand{\ee}{\end{equation}}
\newcommand{\bs}{\begin{split}}
\newcommand{\es}{\end{split}}

\newcommand{\EE}[1]{\langle #1\rangle}
\newcommand{\E}[2]{\left\langle #2\right\rangle_{#1}}
\newcommand{\ds}{\displaystyle}
\newcommand{\Z}{Z}
\newcommand{\floor}[1]{\lfloor#1\rfloor}

\newcommand \rhobar{\bar\rho}
\newcommand \Tr{\mathop{\rm Tr}}

\begin{document}

%% \title[TASEP with Semi-Permeable Boundaries]{On the 
%% Asymmetric Exclusion Process with Semi-Permeable Boundaries}
%% \author{Arvind Ayyer}
%% \address{Arvind Ayyer\\
%% Department of Physics\\
%% Rutgers University\\
%% 136 Frelinghuysen Rd\\
%% Piscataway, NJ 08854.}
%% \email{ayyer@physics.rutgers.edu}
%% \author{Joel L. Lebowitz}
%% \address{Joel L. Lebowitz\\
%% Department of Mathematics \\
%% Rutgers University \\
%% 110 Frelinghuysen Road \\
%% Piscataway, NJ 08854, USA}
%% \email{lebowitz@math.rutgers.edu}
%% \author{Eugene R. Speer}
%% \address{E. R. Speer\\
%% Department of Mathematics \\
%% Rutgers University \\
%% 110 Frelinghuysen Road \\
%% Piscataway, NJ 08854, USA}
%% \email{speer@math.rutgers.edu}

\title[Two Species TASEP with Semi-Permeable Boundaries]{On the Two Species
Asymmetric Exclusion Process with Semi-Permeable Boundaries}
\author[Arvind Ayyer]{Arvind Ayyer$^1$}
\email{ayyer@physics.rutgers.edu}
\author[Joel L. Lebowitz]{Joel L. Lebowitz$^{1,2}$}
\email{lebowitz@math.rutgers.edu}
\author[Eugene R. Speer]{Eugene R. Speer$^2$}
\email{speer@math.rutgers.edu}
\thanks{\hspace{-\parindent}1.\hspace{2.9pt}Department of Physics, 
Rutgers University,
136 Frelinghuysen Rd, Piscataway, NJ 08854.\\
2.\hspace{2.9pt}Department of Mathematics, Rutgers University,
110 Frelinghuysen Rd, Piscataway, NJ 08854.}

\begin{abstract} We investigate the structure of the nonequilibrium
stationary state (NESS) of a system of first and second class particles, as
well as vacancies (holes), on $L$ sites of a one-dimensional lattice in
contact with first class particle reservoirs at the boundary sites; these
particles can enter at site 1, when it is vacant, with rate $\alpha$, and
exit from site $L$ with rate $\beta$.  Second class particles can neither
enter nor leave the system, so the boundaries are {\it semi-permeable}. The
internal dynamics are described by the usual totally asymmetric exclusion
process (TASEP) with second class particles. An exact solution of the NESS
was found by Arita.  Here we describe two consequences of the fact that the
flux of second class particles is zero. First, there exist (pinned and
unpinned) fat shocks which determine the general structure of the phase
diagram and of the local measures; the latter describe the microscopic
structure of the system at different macroscopic points (in the limit
$L\to\infty$) in terms of superpositions of extremal measures of the
infinite system. Second, the distribution of second class particles is
given by an equilibrium ensemble in fixed volume, or equivalently but more
simply by a pressure ensemble, in which the pair potential between
neighboring particles grows logarithmically with distance.  We also point
out an unexpected feature in the microscopic structure of the NESS for
finite $L$: if there are $n$ second class particles in the system then the
distribution of first class particles (respectively holes) on the first
(respectively last) $n$ sites is exchangeable.  \end{abstract}

\maketitle

\section{Introduction} \label{sec:model} 

In recent work Arita \cite{arita0,arita}, using a matrix ansatz, found the
nonequilibrium stationary state (NESS) of a new version of the widely
studied one-dimensional totally asymmetric exclusion process (TASEP)
\cite{spitz}--\cite{sch} (see in particular \cite{be} for
a recent review of matrix methods for the TASEP).  
The model is defined on a subset of the one
dimensional lattice $\bZ$ consisting of $L$ sites. Each site $i$,
$i=1, \dots ,L$, may be occupied by a first class particle, occupied by a
second class particle, or vacant; vacant sites are also referred to as
holes, and first class particles simply as particles.  We shall let these
three possible states correspond to the values $1$, $2$, and $0$,
respectively, of a random variable $\tau_i$; we also introduce the
indicator random variables $\eta_a(i)$, $a=0,1,2$, such that $\eta_a(i)=1$
if $\tau_i=a$ and $\eta_a(i)=0$ otherwise. 

 The internal (bulk) dynamics of the system are given by the usual rules
for the TASEP with second class particles \cite{djls}.  The occupation
variable $\tau_i$ at site $i, i=1,\dots,L-1$, attempts when $\tau_i=1$ or 2
to exchange at rate 1 with $\tau_{i+1}$; when $\tau_i=1$ the exchange
succeeds iff $\tau_{i+1}=0$ or 2, while for $\tau_i=2$ it only succeeds if
$\tau_{i+1}=0$. In other words, a first class particle at site $i$ jumps to
the right by exchanging with either a hole or second class particle at site
$i+1$, while a second class particle can only jump if the site on its right
is empty. At site $i=1$, first class particles enter the system at rate
$\alpha$ provided that site $1$ is vacant ($\tau_1=0$); at site $i=L$,
first class particles leave the system at rate $\beta$ provided that site
$L$ is occupied by a first class particle ($\tau_L=1$).  Second class
particles are thus trapped inside the system; since only first class
particles can cross the boundaries, we refer to these as {\it
semi-permeable}.  (A similar system, but with a different form of
semipermeable boundary, was considered in \cite{kjs}).  An equivalent
system is obtained by interchanging first class particles with holes, left
with right, and $\alpha$ with $\beta$, and this symmetry will be reflected
in the structure of the NESS. The latter will be determined by the
parameters $\alpha, \beta$ and the density $\gamma = n/L$ of second class
particles, where $n$ is the number of second class particles in the system.

\begin{figure}[ht!]
\vspace{-1cm}
\includegraphics[width=10cm]{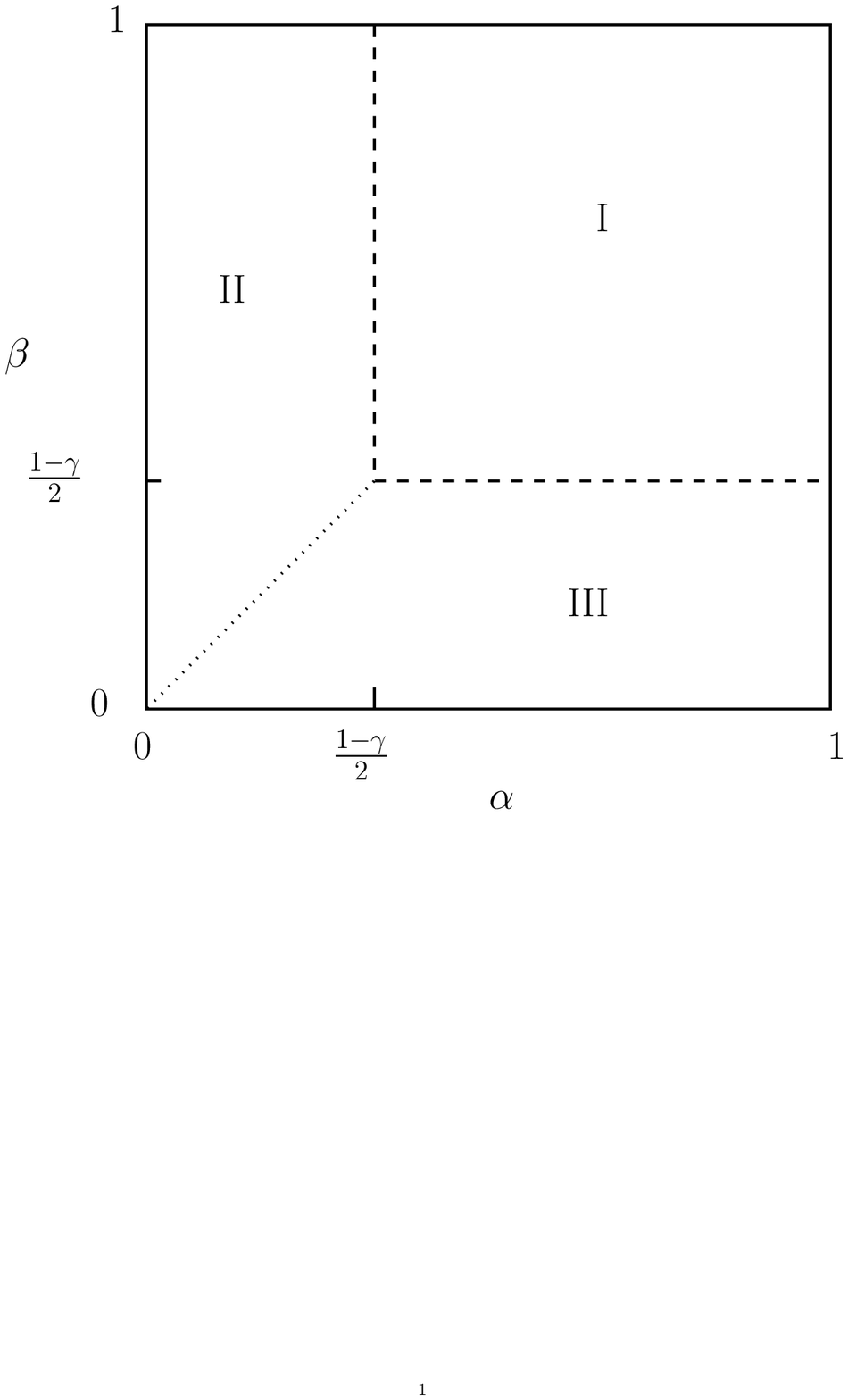} 
\vspace{-5cm}
\caption{The cross section of the phase diagram at a fixed $\gamma$.} 
  \label{fig:phase}
\end{figure}

The phase diagram of this system in the limit $L \to \infty$ is given in
Figure \ref{fig:phase} \cite{arita}.  The diagram is determined by the
distinct formulas for the (first class) particle current $J_1$ in the
different regions:
\be
J_1 = \left\{ \begin{array}{lll}
\displaystyle \frac{1-\gamma^2}{4}, 
 & \quad \text{for }\alpha,\beta \geq \alpha_c \;\ (\text{region I}),\\
\\
\displaystyle \alpha(1-\alpha), 
  & \quad\text{for } \alpha< \alpha_c,\alpha<\beta \;\ (\text{region II}),\\
\\
\displaystyle \beta(1-\beta), 
  &  \quad\text{for } \beta <\alpha_c,\beta<\alpha \;\ (\text{region III});
\end{array} \right.\label{Jdef}
\ee
 here the critical value $\alpha_c$ of $\alpha$ and $\beta$ is
\be \label{alphac}
\alpha_c = \frac{1-\gamma}{2}.
\ee
 The current $J_2$ of the (trapped) second class particles must vanish and
the current $J_0$ of holes satisfies $J_0=-J_1$.  Note that the form of
$J_1$ as a function of the parameters characterizes the phase plane
regions, and that $J_1$ is continuous but not smooth across all the
boundaries.

The phase diagram is similar to that of the open one component TASEP
\cite{dehp,sd} and indeed in the limit $\gamma\to0$ reduces to it; moreover,
in regions~II and III the current is independent of $\gamma$ and takes the
same values as in the one component case, although the size of these
regions shrinks as $\gamma$ increases.  As we discuss in
Section~\ref{sec:localbdries}, however, there will be residual differences
between the local microscopic states of the one species model and
$\gamma\to0$ limit of the two species model; in particular, there remain an
infinite number of second class particles near one or both boundaries of
the two species system.  Note also that there is a discontinuity, equal to
$\gamma$, in the derivative of $J_1$ with respect to $\alpha$ ($\beta$) on
the I/II (I/III) boundary; one might say that the order of the phase
transition in $J_1$ when $\gamma\ne0$ differs from that when
$\gamma=0$.

The macroscopic density profiles $\rho_a(x)$ in the
NESS, $a=0,1,2$, defined by 
 \be \label{rhodef}
\rho_a(x)=\lim_{L\to\infty,\,n/L\to\gamma,\,i/L\to x}\EE{\eta_a(i)},\qquad
   0\le x\le1,
 \ee
 with $\EE{\cdot}$ the expected value in the NESS, have been computed in
\cite{arita}; the results are summarized in Table~\ref{table:dens} (but see
Remark~\ref{rem:nonunique} below).  Knowing any two of these densities
determines the third, via $\sum_a\rho_a(x)=1$.  In fact, knowing
$\rho_1(x)$ for all $x$ and all values of $\alpha$ and $\beta$ determines
$\rho_0(x)$, from the particle-hole symmetry, and hence all profiles, but
for clarity we give in Table \ref{table:dens} both $\rho_1(x)$ and
$\rho_0(x)$.  In region II the system divides itself into two parts,
$x<x_0$ and $x>x_0$, with different formulas for $\rho_0(x)$; similarly in
region III there are different formulas for $\rho_1(x)$ for $x<x_1$ and
$x>x_1$.  Here
 \be \label{x0x1}
  \begin{split}
  x_0&=1-\frac{\gamma}{1-2\alpha},\qquad 
    \alpha\le\beta,\quad\alpha<\alpha_c;\\
 x_1&=\frac{\gamma}{1-2\beta},\qquad 
     \beta\le\alpha,\quad \beta<\alpha_c.
 \end{split}
 \ee
On the II/III boundary $\alpha=\beta<\alpha_c$, the {\it shock line}, the
profiles include linear regions:
 \be \label{linear}
 \begin{split}
\rho_0(x) &= \begin{cases}
  \ds\frac{x_0-x}{x_0}(1-\alpha)+\frac{x}{x_0}\,\alpha,& 0\le x\le x_0,\\
   \alpha,& x_0\le x\le1\end{cases}\\
\rho_1(x) &= \begin{cases}
    \alpha,& 0\le x\le x_1,\\
   \ds\frac{1-x}{1-x_1}\,\alpha+\frac{x-x_1}{1-x_1}(1-\alpha),& x_1\le x\le1.
   \end{cases}
 \end{split}
 \ee
 These arise from averaging over the position of a shock, as in the
one species TASEP; further discussion is given below.

 \begin{table}[ht]
\centering
\caption{Density profiles in different regions of the phase plane. Note
  that $x_0$ is defined only in region~II and on its boundaries, and $x_1$
  only in region~III and on its boundaries.}
\begin{tabular}{|l|c|c|c|c|}
\hline
Region &\multicolumn2{c|}{$\rho_1(x)$}&\multicolumn2{c|}{$\rho_0(x)$}  \\
\hline
I  & \multicolumn2{c|}{$\alpha_c$} 
     & \multicolumn2{c|}{$\alpha_c$} \\
I/II boundary &  \multicolumn2{c|}{$\alpha_c$} 
     & \multicolumn2{c|}{$\alpha_c$} \\
I/III boundary &  \multicolumn2{c|}{$\alpha_c$}
     & \multicolumn2{c|}{$\alpha_c$} \\
\hline & $x < x_1$ &  $x>x_1$
   & $x < x_0$ &  $x>x_0$\\
\hline
II  & \multicolumn2{c|}{$\alpha$} & $1-\alpha$  & $\alpha$ \\
\cline{2-5}
III  & $\beta$  & $1-\beta$ & \multicolumn2{c|}{$\beta$}  \\
\cline{2-5}
II/III boundary (Shock Line) & $\alpha\,(=\beta)$ & linear 
  & linear & $\alpha\,(=\beta)$  \\
\hline
\end{tabular} \label{table:dens} 
\end{table}

\begin{rem}\label{rem:nonunique} (a) The density values $\rho_a(x)$ at the
boundaries $x=0,1$ and at the fixed shocks $x=x_0,x_1$ may depend the way
the limit \eqref{rhodef} is taken.  The boundary cases $x=0,1$ were
discussed in \cite{arita} except on the I/II and I/III boundaries.  We
discuss the limits at $x_0,x_1$ in Section~\ref{sec:localbulk} and
\ref{sec:localbdries}; this gives some further information about limits at
the boundaries since $x_0=0$ and $x_1=1$ on the I/II and I/III boundaries,
respectively.

 \smallskip\noindent
 (b) In the one-component model the phase plane regions corresponding to
I, II, and III are called the maximum current, low density, and
high density regions, respectively.  We do not adopt that terminology here
since in region III the particle density is low for $x<x_1$.
\end{rem} 

\begin{figure}[ht!]
\includegraphics[width=6.2cm]{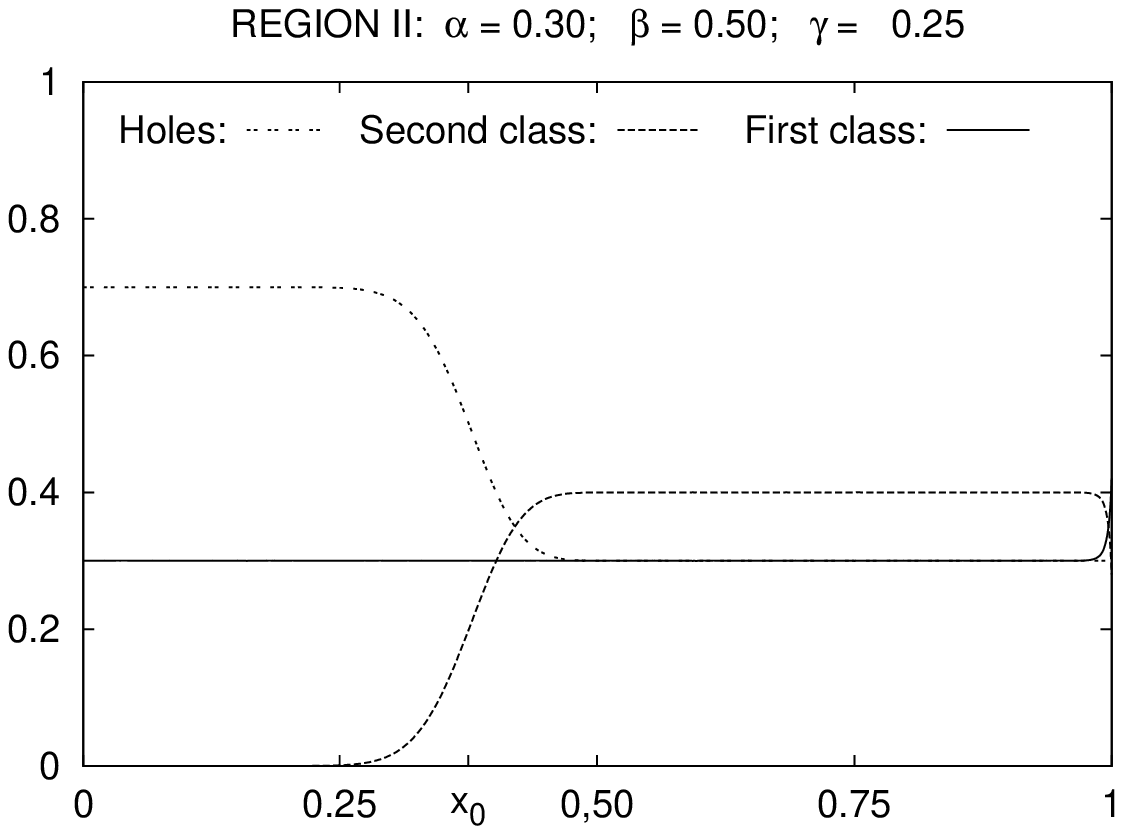}
\includegraphics[width=6.2cm]{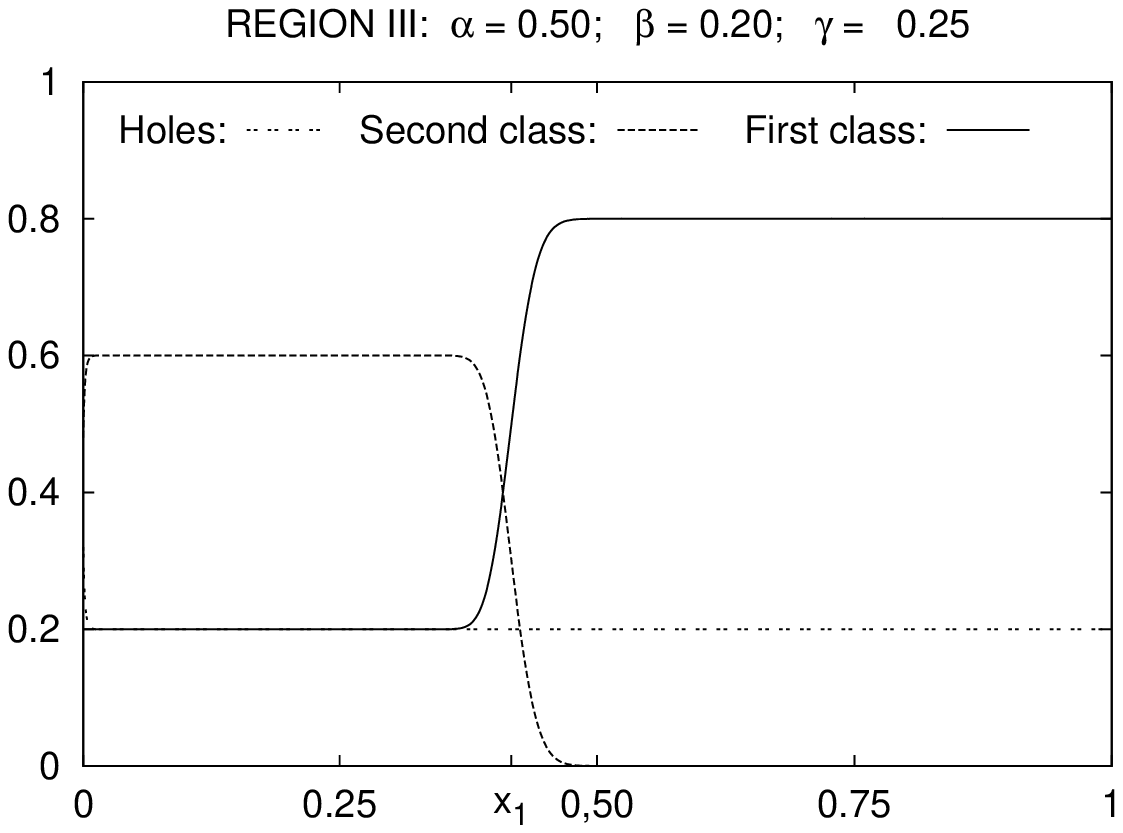}

\hbox to\hsize{\hfil(a)\hfil\hfil(b)\hfil}

\bigskip
\includegraphics[width=6.2cm]{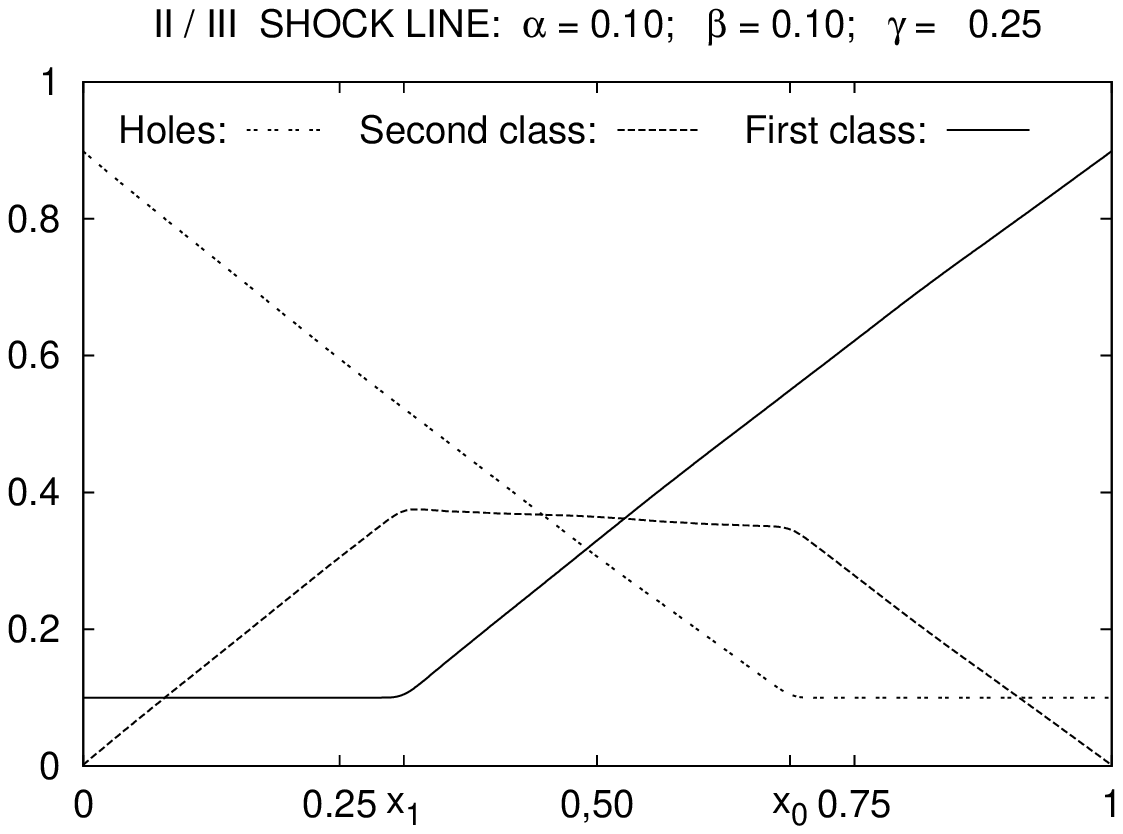}
\includegraphics[width=6.2cm]{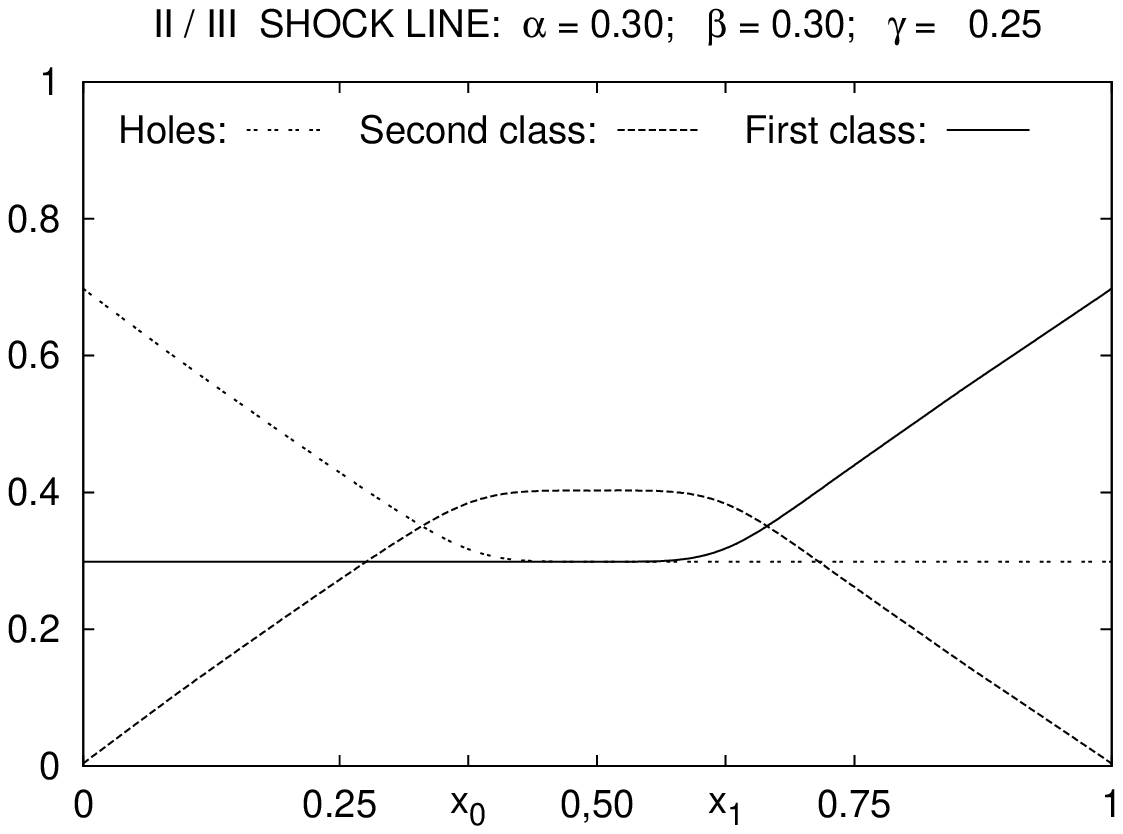}

\hbox to\hsize{\hfil(c)\hfil\hfil(d)\hfil}

\bigskip
\includegraphics[width=6.2cm]{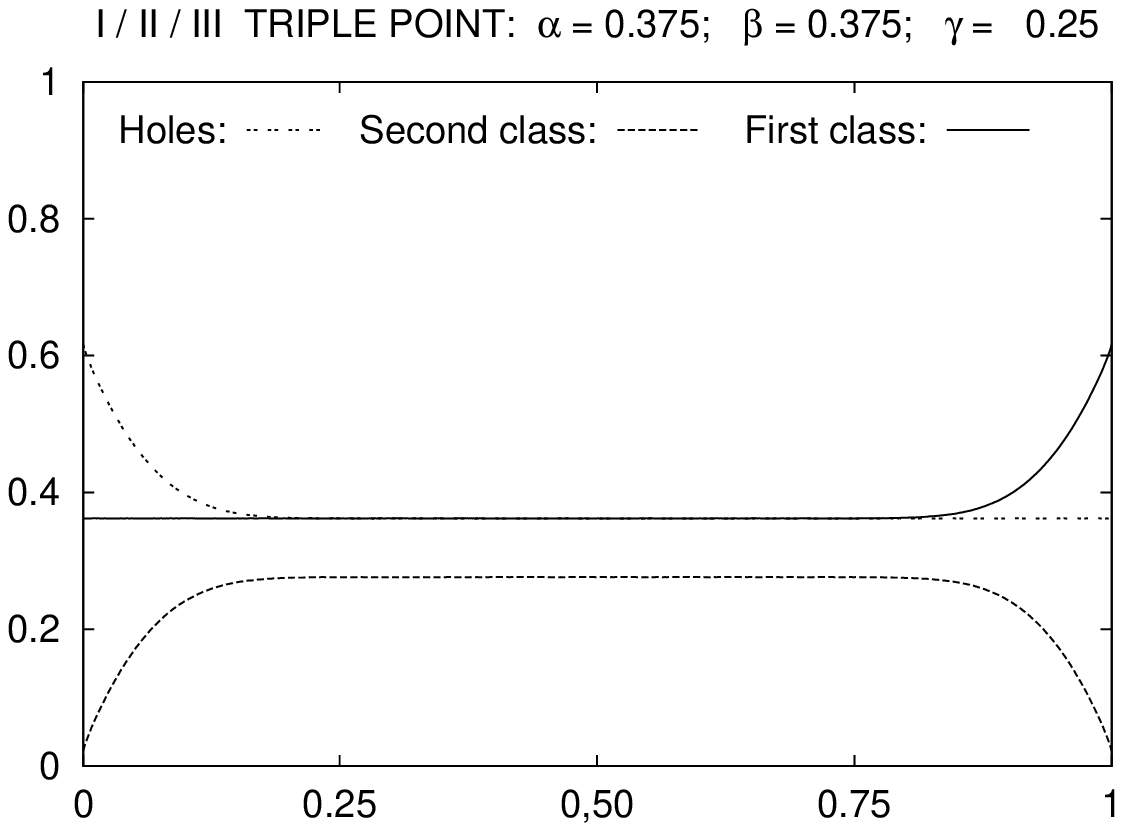}
\includegraphics[width=6.2cm]{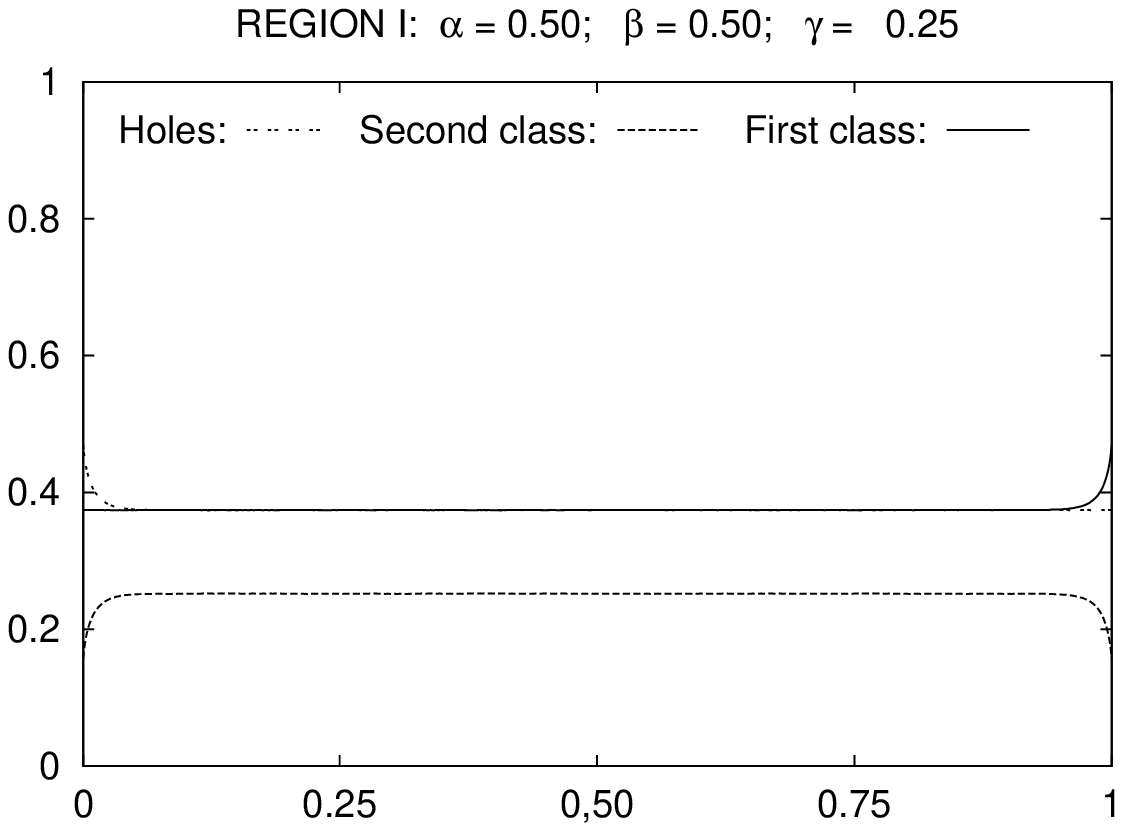}

\hbox to\hsize{\hfil(e)\hfil\hfil(f)\hfil}

\caption{Density profiles in a system with $L=1000$.} 
\label{fig:profs}
\end{figure}

Some typical profiles, obtained from simulations, are shown in
Figure~\ref{fig:profs}.  Since these are from a finite system, they do not
coincide perfectly with the description in Table~\ref{table:dens}: there
are boundary effects, and the density transitions in regions~II and III, at
$x_1$ and $x_0$ respectively, have nonzero width of order $\sqrt{L}$.  This
is related to the nonuniqueness of the limit \eqref{rhodef} at these
points, as mentioned above, and is discussed in
Section~\ref{sec:localbulk}.

We now give an intuitive discussion of some of the phenomena that give rise
to these profiles.  Consider a macroscopically uniform portion of our
system in the limit $L \to \infty$, with densities of holes, particles and
second class particles denoted by $\rho_0$, $\rho_1$ and $\rho_2$,
respectively, where $\rho_0+\rho_1+\rho_2=1$.  In such a region
the measure will be the known translation invariant measure, with these
densities, for the two species TASEP (see \cite{djls,speer,ffk} and 
the discussion in Section~\ref{sec:con}); in this measure
(which is not a product measure) the first class particles considered
separately, and the holes considered separately, are distributed according
to product measures, so that $J_1=\rho_1 (1-\rho_1)$ and
$J_0=-\rho_0(1-\rho_0)$.  Thus from $J_0=-J_1$ it follows that in any
uniform stretch of the NESS either
 \be \label{alt}
\rho_1=\rho_0=(1-\rho_2)/2 \qquad \text{or} \qquad \rho_2=0,\ \rho_1 = 1-
\rho_0. 
 \ee
 This fact, which may be seen in the results of \cite{arita}, is key to
understanding the gross structure of the densities in different regions of
the phase diagram.

These density profiles differ from those the single-species open TASEP in
two notable ways: in regions II and III the density profiles have a point
of discontinuity, and on the II/III boundary the linear region occupies
only part of the system.  These and other properties can be understood in
terms of the occurrence of a {\em fat shock}.  By this term we refer, not to a
broadening of the sharp shock usually seen in the TASEP (as can occur in
the partially asymmetric model \cite{dls1}), but rather to a macroscopically
uniform interval which contains all the second-class particles, and thus
conforms to the first alternative in \eqref{alt}.  We may think of this fat
shock as bounded by shocks of the usual sort occurring in two different one
species TASEP systems which are obtained by making appropriate
identifications of second class particles with either first class particles
or holes.  To see how this occurs, recall that if one either identifies
first and second class particles by coloring holes black and both kinds of
particles white, or else identifies second-class particles and holes by
coloring these species red and first-class particles blue, then the
black/white particles, as well as the red/blue particles, form standard two
species TASEPs in the bulk.  The dynamics at the boundaries is different,
since some white ``particles'' or red ``holes'' will be trapped in the
system.  A careful justification of the conclusions below is given in
Sections~\ref{sec:fat} and \ref{sec:localbulk}.

\begin{figure}[ht!]
\hspace{10cm}
\vspace{-4cm}
\includegraphics[width=10cm]{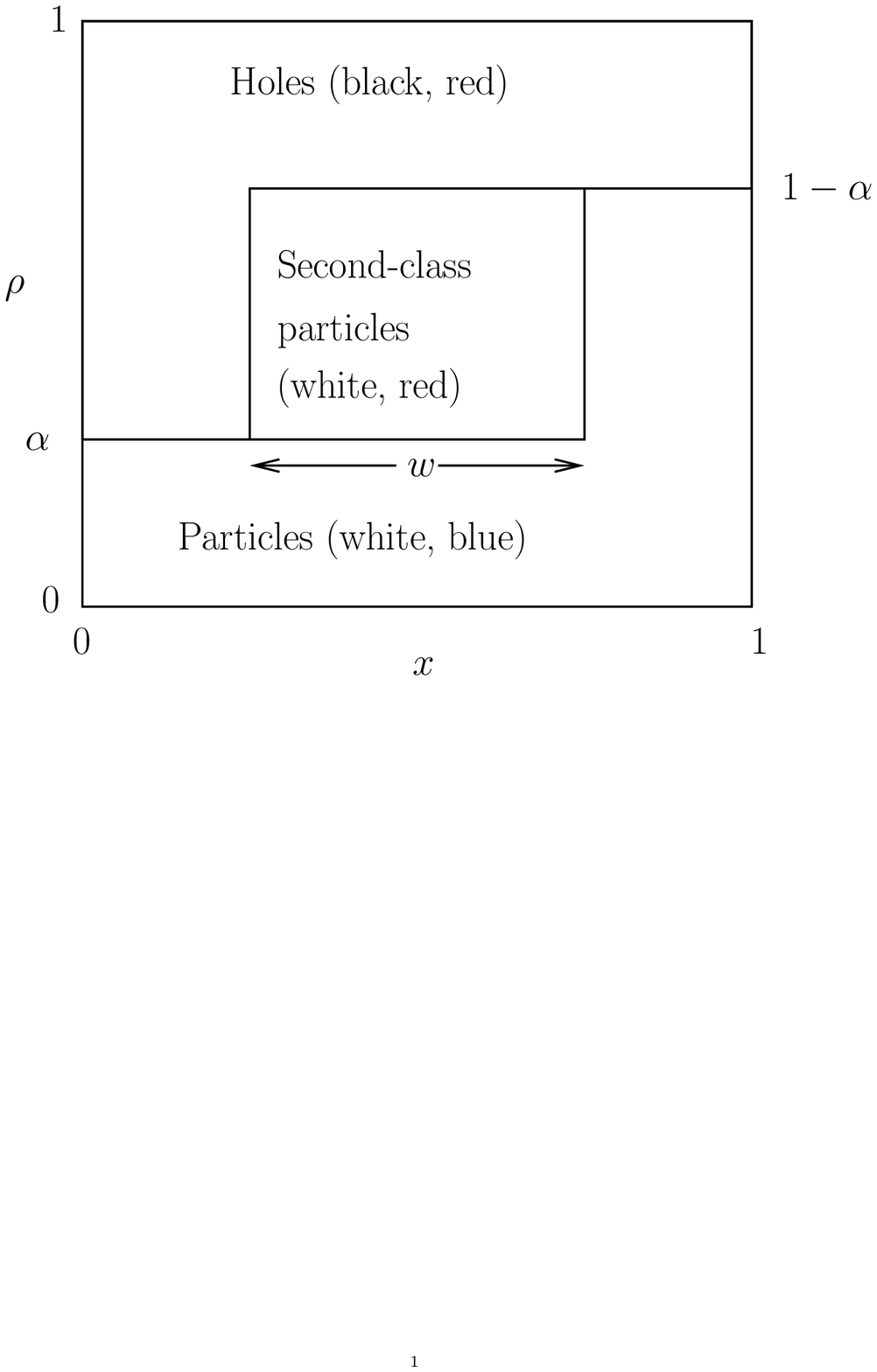}
\vspace{-2cm}
\caption{Shock interpretation at $\alpha=\beta<\alpha_c$.  Densities 
  $\rho_a(x)$ are plotted against $x$, with the convention that at the
  height at $x$  of the region labeled with particle type $a$ is
  $\rho_a(x)$.  The fat shock may in fact be located anywhere in the system.} 
\label{fig:shock}
\end{figure}

Consider now the behavior of the system on the boundary of regions II and
III (the shock line).  Then by previous analysis, see e.g. \cite{ligg2}, one
knows that a typical profile for the one species model contains a
shock between a region of density $\alpha$ on the left and $1-\alpha$ on
the right; the shock position has mean velocity zero and its (fluctuating)
position is uniformly distributed over the system.  We see this same
behavior for both the black/white and red/blue systems described above,
with the black/white shock necessarily located to the left of the red/blue
one.  The {\it typical} profile at any given time looks on the macroscopic
scale like Figure~\ref{fig:shock}, where the convention is that at any
point $x$ the height of the region labeled with particle type $a$ is
$\rho_a(x)$.

Clearly both shock fronts have mean velocity zero and are trapped in the
system, and since the total number of second-class particles is $\gamma L$,
the macroscopic width $w$ of the fat shock must satisfy
$w=\gamma/(1-2\alpha)$.  This forces the two shock fronts to move (i.e.,
fluctuate) in collusion so as to keep the macroscopic shock width fixed; we
expect this fluctuation, as for the shock in the single component TASEP on
its shock line, to be on a diffusive time scale growing as $L^2$.  The
density profiles $\rho_a(x)$ arise as averages over the shock position, and
this gives rise to the linear profiles \eqref{linear}; in contrast to the
situation in the one species case, however, here they occupy only part of
the system because the shock can fluctuate only over an interval of width
$1-w$.  The shock fluctuation is also reflected in the structure of the
local measures obtained in the limit $L\to\infty$, which are superpositions
of states with different densities (see Section~\ref{sec:localbulk}).  The
critical value of $\alpha$ occurs when the fat shock fills the system,
i.e., when $w=1$, from which we regain \eqref{alphac}.

The situation in regions II and III is similar. The fat shock width is in
general
 \begin{equation} \label{wdef}
w(\alpha,\beta,\gamma)=\frac{\gamma}{1-2\alpha\wedge\beta\\}, 
 \end{equation}
  where $\alpha\wedge\beta=\min\{\alpha,\beta\}$; in region II the shock is
pinned to the right boundary, and in region III is pinned to the left
boundary.  Since the shock is fixed it gives rise to discontinuities in the
density profiles; see Figure~\ref{fig:profs} as well as the discussion of a
related model, where similar behavior occurs, in Section~\ref{sec:con}
(Figure~\ref{fig:semiprofs}).  There is no corresponding discontinuity in
the single-species TASEP ($\gamma=0$) because in that case there is a
single shock of zero macroscopic width.  In region I, $w$ as given in
\eqref{wdef} is greater than 1 and the fat shock fills the system; the
density profiles are uniform and conform to the first alternative of
\eqref{alt}.

The outline of the rest of the paper is as follows.  In
Section~\ref{sec:matrix} we discuss the matrix method for this system.
We use a different representation of the matrices from that of
\cite{arita}, which makes it easier to prove certain features of the
NESS discussed later.  In Section~\ref{sec:exch} we show that the
marginal distribution induced by the NESS on particles in the first
$n$ states of the system, and on holes in the last $n$, is
exchangeable, i.e., that the probability of finding $r$ first class
particles (holes) on some specified set of $r$ sites among the first
(last) $n$ is independent of the choice of sites.  In
Section~\ref{sec:fat} we establish the fat shock picture described
above.  In Section~\ref{sec:localbulk} determine the local measures,
in the bulk, for the infinite volume limit of the system, and in
Section~\ref{sec:localbdries} consider the local measures near the
boundaries, focusing on a Bernoulli property which is a consequence of
the exchangeability established in Section~\ref{sec:exch}.  In
Section~\ref{sec:equil} we show that the second class particles form
an equilibrium system, most simply described by a pressure
ensemble. This is related, in our case, to the fact that the current
of second class particles is zero. For similar situations, see
\cite{rps,bdgjl2}.

In Section~\ref{sec:con} we make some concluding remarks and, in particular,
describe some closely related models. One such model is a generalization of
the standard ``defect particle'' model; another describes a system of first
class particles, second class particles, and  holes on a ring with one
semi-permeable bond which second class particles cannot cross.  Several more
technical remarks are recorded in the appendices.

\section{The matrix ansatz} \label{sec:matrix}

 The stochastic system described in Section~\ref{sec:model} is ergodic in
finite volume $L$ and thus there exists a unique invariant measure
$\mu_{L,n}^{\alpha,\beta}$ on the configuration space
 \be
Y_{L,n}\equiv \{\, (\tau_1,\ldots,\tau_L)\mid\tau_i=0,1,2;\ \tau_i=2
 \text{ for $n$ values of $i$}\,\},
 \ee
 where from now on we will assume that $0<n<L$.  This measure may be
obtained from a matrix ansatz \cite{arita}, combining the matrix algebra of
\cite{djls} (which discussed the system with the same constituents as in
the current work, but on a ring) with the treatment of the one species open
system via matrix-elements from \cite{dehp}.  One introduces matrices
$X_0$, $X_1$, and $X_2$ and vectors $|V_\beta \rangle$ and
$\langle W_\alpha |$ which satisfy
\be \label{dereq}
X_1 X_0  =  X_1+X_0, \qquad
X_1 X_2  =  X_2, \qquad
X_2 X_0  =  X_2 ,
\ee
and
\be \label{deric}
\langle W_\alpha | X_0  =  \frac{1}{\alpha} \langle W_\alpha|, \qquad
X_1 |V_\beta \rangle  =  \frac{1}{\beta} |V_\beta \rangle.
\ee
Then for a configuration $\tau = (\tau_1,\cdots,\tau_L)\in Y_{L,n}$ the
probability of $\tau$ in the invariant measure is
\be \label{probtau}
\E{\mu^{\alpha,\beta}_{L,n}}{\tau} 
  =Z^{\alpha,\beta}(L,n)^{-1}
 \langle W_\alpha| X_{\tau_1} \cdots X_{\tau_L} |V_\beta \rangle,
\ee
where $Z^{\alpha,\beta}(L,n)$ is the normalization factor
 \begin{equation}\label{Zdef}
Z^{\alpha,\beta}(L,n)
  =\sum_{\tau\in Y_{L,n}}
   \langle W_\alpha| X_{\tau_1} \cdots X_{\tau_L} |V_\beta \rangle,
 \end{equation}
which, with a slight misuse of the nomenclature of equilibrium statistical
mechanics, we call the {\it partition function}.  We will frequently omit
superscripts such as $\alpha,\beta$ in \eqref{Zdef} when no confusion can
arise.
 
 We will work throughout in a realization of
\eqref{dereq}--\eqref{deric}, different from that of \cite{arita}, for
which the matrices and vectors have the further properties
 \begin{gather}\label{conseq}
 X_2=X_1X_0-X_0X_1=|V_1\rangle\langle W_1|, \qquad
 X_2|V_\beta\rangle = |V_1\rangle,\qquad
 \langle W_\alpha|X_2=\langle W_1|,\\
   \langle W_\alpha|V_1\rangle =    \langle W_1|V_\beta\rangle = 1 \qquad
   \text{for all $\alpha,\beta$,}\label{conseq2}
 \end{gather}
 where $|V_1\rangle$ and $\langle W_1|$ are vectors satisfying
\eqref{deric}.  Note that $X_2$ is then a one-dimensional projection
operator.  The realization is given in Appendix~\ref{app:rep}, but we will
need no consequences beyond \eqref{conseq} and \eqref{conseq2}.  Because of
\eqref{conseq2} we make the convention that
$Z^{\alpha,1}(0,0)=Z^{1,\beta}(0,0)=1$.

\begin{rem} \label{finite} The nature of $X_2$ in this representation shows
that certain  distributions obtained from
\eqref{probtau} factorize.  Let
$Q_1,\ldots,Q_n$ denote the (random) positions of the second class
particles in the system (note that these can be ordered once and for all).
Then the probability that the $j_1^{\rm th}$, $j_2^{\rm th}$, \dots,
$j_m^{\rm th}$ second class particles are located on sites
$q_{j_1},\ldots,q_{j_m}$ is
 \begin{align}
 &\hskip-20pt \mu^{\alpha,\beta}_{L,n}(Q_{j_i}=q_{j_i},i=1,\ldots,m)
 =  Z^{\alpha,\beta}(L,n)^{-1}
   Z^{\alpha,1}(q_{j_1}-1,j_1-1)\nonumber\\
 &\hskip10pt \times \prod_{i=2}^{m}Z^{1,1}(q_{j_i}-q_{j_{i-1}}-1,j_i-j_{i-1}-1)
      \; Z^{1,\beta}(L-q_{j_m},n-j_m).\label{factorZ}
 \end{align}
 Moreover, if we condition on the event that $Q_j=q$, i.e., that the
$j^{\rm th}$ second class particle is located at site $q$, then the
conditional measure is a product of the measures associated with the sites
before and after $j$, so that if $\tau$ is a configuration consistent with
this event then
 \be\label{factortau}
\mu_{L,n}^{\alpha,\beta}(\tau_1,\ldots,\tau_L\mid Q_j=q)
  =\mu_{q-1,j-1}^{\alpha,1}(\tau_1,\ldots,\tau_{q-1})
    \mu_{L-q,n-j}^{1,\beta}(\tau_{q+1},\ldots,\tau_L).
 \ee
 A factorization property of this type is also known for the translation
invariant measures for the two species TASEP \cite{djls,ffk}.  Similar
expressions are easily obtained when conditioning on the presence of
several specified second class particles at specified sites.  We will use
\eqref{factorZ} and \eqref{factortau} in Sections~\ref{sec:fat} and
\ref{sec:localbulk}, when we discuss the fat shock and describe local
measures in the NESS.  \end{rem}

\section{Exchangeability of Measures} \label{sec:exch}

In this section we demonstrate a remarkable property of the finite-volume
NESS with $n$ second class particles: the {\it exchangeability}
\cite{feller} of the measure on first class particles within the first $n$
sites, or equivalently on holes in the last $n$ sites.  Specifically, this
means that for any $r\le n$ the probability of finding first class
particles on the $r$ sites $1 \leq i_1 < i_2 < \dots < i_r \leq n$ depends
only on $r$, i.e., is independent of the choice of positions
$i_1,i_2,\dots,i_r$.  When $r=1$ this is implicit in (38) of \cite{arita},
although it is not emphasized.  As a consequence of the ideas of the proof
we will also obtain, for any $i,j$ with $i,j\ge1$ and $i+j-1\le L$, the
probability of finding a block of $j$ consecutive first class particles
starting at site $i$; this generalizes the density formula of \cite{arita},
which corresponds to $j=1$.

The key quantity for our arguments is the probability of finding (first
class) particles at sites $i_1,\dots,i_{r-1}$ together with a block of $j$
particles starting at site $i_r$, where $r\ge1$, $i_1<\cdots<i_{r-1}$, and
$i_r>i_{r-1}+1$; we allow $j=0$, with the interpretation that in this case
there is no restriction on what happens at site $i_r$ or any succeeding
sites.  Thus the probability in question is
$Z(L,n)^{-1}E_r(L,n;i_1,\ldots,i_r;j)$, where $E_r$ is a sum of weights for
certain configurations $\tau\in Y_{L,n}$:
\be\label{E}
E_r(L,n;i_1,\dots,i_r;j) = \hskip-30pt\sum_{\substack{
\tau_{i_1}=\dots=\tau_{i_{r-1}}=1\text{ if $r\ge2$}\\
 \tau_{i_r}=\tau_{i_r+1}=\dots=\tau_{i_r+j-1}=1\text{ if $j\ge1$}}}\hskip-30pt
\langle W_\alpha| X_{\tau_1} \dots X_{\tau_L} |V_\beta \rangle
\ee
 In \eqref{E} we must have $i_r+j-1 \leq L$, since there are only $L$
sites, and $r+j-1 \leq L-n$, since there are $n$ second class particles.
For certain parts of the analysis we will have to consider separately two
cases, in which these two inequalities respectively provide the effective
bounds on $j$:

 \smallskip\noindent
 {\bf Case 1:} \quad $i_r\ge n+r$, so that  $0\le j\le L-i_r+1$; 

 \smallskip\noindent
 {\bf Case 2:}  \quad $i_r\le n+r-1$, so that  $0\le j\le L-n-r+1$.

 \smallskip\noindent

 We will analyze the $E_r$ using a simple recursion:
 \be   \label{recur}
E_r(L,n;i_1,\dots,i_r;j) =
E_r(L,n;i_1,\dots,i_r;j+1)+E_r(L-1,n;i_1,\dots,i_r;j-1).
 \ee
 This holds whenever all terms are defined, which requires that $j$ be
positive and satisfy $j\le L-i_r$ in Case~1 and $j\le L-n-r$ in Case~2.  To
derive \eqref{recur} we consider the value of $\tau_{i_r+j}$ in each term
of the sum \eqref{E}.  Terms with $\tau_{i+j}=1$ sum precisely to
$E_r(L,n;i_1,\ldots,i_r;j+1)$, and for terms with $\tau_{i+j}=0$ or
$\tau_{i+j}=2$ we use the matrix algebra to reduce
 \be
X_{\tau_{i_r+j-1}}X_{\tau_{i_r+j}}
  =\begin{cases}X_1X_0=X_1+X_0,& \text{if $\tau_{i_r+j}=0$,}\\
   X_1X_2=X_2,& \text{if $\tau_{i_r+j}=2$;}\end{cases}
 \ee
 the resulting sum is just $E_r(L-1,n;i_1,\dots,i_r;j-1)$.  

To determine the $E_r$ the recursion \eqref{recur} must be supplemented by
boundary conditions at the maximum and minimum values of $j$.  When $j=0$,
\eqref{E} gives
 \be\label{BC1}
E_r(L,n;i_1,\ldots,i_r;0)
   =\begin{cases}E_{r-1}(L,n;i_1,\ldots,i_{r-1};1),& \text{if $r\ge2$},\\
     Z(L,n),& \text{if $r=1$}.\end{cases}
 \ee
 The value of $E_r$ for $j$ maximal is case dependent.  In Case~1, if $j$
takes its maximum possible value $L-i_r+1$ then each matrix product in
\eqref{E} ends with
$X_1^{L-i_r+1}|V_\beta\rangle=\beta^{-(L-i_r+1)}|V_\beta\rangle$, so that
 \begin{align} \nonumber
 E_r(L,n;i_1,\ldots,i_r;L-i_r+1)& \\
 &\hskip-80pt = \begin{cases}
  \beta^{-(L-i_r+1)}E_{r-1}(i_r-1,n;i_1,\ldots,i_{r-1};1),
      & \text{if $r\ge2$},\\
  \beta^{-(L-i_r+1)}Z(i_r-1,n),& \text{if $r=1$},\end{cases}
   \qquad \text{(Case 1).} \label{BC2/a}
 \end{align}
 In Case~2, if $j$ takes its maximum possible value $L-n-r+1$ then the
rightmost factor in the matrix product in \eqref{E} is $X_2$ and there are
no factors of $X_0$; using the matrix algebra relations $X_1X_2 = X_2$,
$X_2^2 = X_2$ and $X_2 |V_\beta \rangle = |V_1\rangle$, we have that
 \be \label{BC2/b}
 E_r(L,n;i_1,\ldots,i_r;L-n-r+1) =1 \qquad \text{(Case 2).}  
 \ee

\begin{lem}\label{lem:recur} The recursion \eqref{recur}, together with the
boundary conditions \eqref{BC1} and either \eqref{BC2/a} or \eqref{BC2/b},
determines $E_r(L,n;i_1,\dots,i_r;j)$ by an inductive computation.
\end{lem}  

\begin{proof}The primary induction is on increasing values of $r$, with $n$
held fixed throughout.  The inductive assumption that the $E_{r-1}$ are
known is needed when \eqref{BC1} or \eqref{BC2/a} is applied, and since for
$r=1$ the right hand side of these equations is a (known) partition
function one may treat all values $r\ge1$ uniformly.  Then for fixed $r$
($r\ge1$) and $i_1,\ldots,i_r$ we induce on increasing values of $L$: for
the minimum possible value, $L=i_r$, we must be in Case 1 and have $j=0$ or
$j=1$, so that all $E_r(L,n;i_1,\ldots,i_r;j)$ are determined by
\eqref{BC1} and \eqref{BC2/a}.  For any larger value of $L$ either
\eqref{BC2/a} or \eqref{BC2/b} determines $E_r(L,n;i_1,\ldots,i_r;j)$ for
the maximum possible value of $j$ and we may then induce downward on $j$
using \eqref{recur}.  \end{proof}

We can now verify  exchangeability; we show that
$E_r(L,n;i_1,\dots,i_r;j)$ is equal to the corresponding weight with the
sites $i_1,\dots,i_r$ in standard positions $1,\ldots,r$.

\begin{thm} \label{thm:exch}
Let $1\leq i_1 < i_2 <\dots<i_r$ be sites such that $i_r \leq n+r-1$, 
and let $j$ be an integer less than or equal to $L-n-r+1$. Then
\be \label{excheq}
E_r(L,n;i_1,\dots,i_r;j) = E_1(L,n;1;r+j-1).
\ee

\end{thm}

\begin{proof} The proof is by induction on $r$.   By
Lemma~\ref{lem:recur} it suffices to show that $E_1(L,n;1;r+j-1)$ satisfies
the same recurrence relation \eqref{recur} and boundary conditions
\eqref{BC1}, \eqref{BC2/b} as does $E_r(L,n;i_1,\dots,i_r;j)$.  This
follows immediately from  the corresponding relations for $E_1$ and, for
$r\ge2$, the induction assumption.
\end{proof}

   We finally give the explicit formula for $E_1$ which, by \eqref{excheq},
also provides a formula for $E_r(L,n;i_1,\dots,i_r;j)$ when $i_r\le n+r-1$.
This result will not be needed in the remainder of the paper.

The formula involves the Catalan triangle numbers \cite{cattri}
\be \label{eqcattri}
C^{m}_{n} = \binom{m+n}{n} \frac{m-n+1}{m+1},
\ee
 which satisfy the recursion 
\be \label{C1}
C^{m-1}_n + C^m_{n-1} = C^m_n
\ee
 and the boundary conditions
\be \label{C2}
C^m_{-1}=0, \qquad C^m_0 = 1, 
\qquad C^m_m = C^m_{m-1} = \frac1{m+1}\binom{2m}{m}.
\ee
 We then define the additional  constants
 \begin{align}\label{c}
c_{j,k}
  &=\begin{cases}C^{k-1}_{k-j},& \text{if $j\ge1$,}\\
 %   \noalign{\smallskip}
 \delta_{k,0},& \text{if  $j=0$;}\end{cases}\\
\label{d} % \noalign{\medskip}
d_{m,j,k} 
 &=\begin{cases}
  \binom{m-j+k}{k}-\binom{m-j+k}{k-j},& \text{if $j\le m$,}\\
%   \noalign{\smallskip}
  \delta_{k,0},& \text{if $j=m+1$.}\end{cases}
\end{align}
 In \eqref{c}--\eqref{d} the  convention is that $\binom{p}{q}=0$ for integer
$p,q$ with $p\ge0$, $q<0$.  

\begin{thm} \label{thm:dens}
{\bf Case 1:} If $n+1\le i\le L$ and $0\le j\le L+1-i$ then
 \be   \label{F}
E_1(L,n;i;j)=\sum_{k=j}^{L-i}c_{j,k}Z(L-k,n)
  + Z(i-1,n)\sum_{k=0}^{L-i-1}d_{L-i,j,k}\beta^{-L+i+k-1}.
 \ee
{\bf Case 2:} If  $1\le i \le n$ and $0\le j\le L-n$ then  
 \be   \label{G}
E_1(L,n;i;j)=\sum_{k=j}^{L-n}c_{j,k}Z(L-k,n),
 \ee
\end{thm}

\begin{proof}
 {\bf Case 1:} We temporarily denote the right hand side of \eqref{F}
by $F(L,n;i;j)$.  By Lemma~\ref{lem:recur} it suffices to verify that
$F$ satisfies relations corresponding to \eqref{BC1}, \eqref{BC2/a}, and
\eqref{recur}.  It will be convenient to denote the two terms in \eqref{F}
by $F_1(L,n;i;j)$ and $F_2(L,n;i
;j)$, respectively.

Since $d_{L-i,0,k}=0$ and $c_{0,k}=\delta_{k,0}$ we have immediately that
$F(L,n;i;0)=Z(L,n)$ (compare \eqref{BC1}).  Moreover, the sum defining
$F_1(L,n;i;L-i+1)$ is empty and so from $d_{L-i,L-i+1,k}=\delta_{k,0}$ we
have $F(L,n;i;L-i+1)=\beta^{-L+i-1}Z(i-1,n)$ (compare \eqref{BC2/a}).
It remains to check the equivalent of \eqref{recur}, which we shall show is
satisfied by $F_1$ and $F_2$ separately; recall that in \eqref{recur},
$1\le j\le L-i$.  For $j\ge2$,
 \begin{align}
 F_1(L,n;i;j+1) +  F_1(L-1,n;i;j-1)&\nonumber\\
 &\hskip-170pt = \sum_{k=j+1}^{L-i}C^{k-1}_{k-j-1}Z(L-k,n)+
    \sum_{k=j-1}^{L-1-i}C^{k-1}_{k-j+1}Z(L-1-k,n)\nonumber\\
 &\hskip-170pt = \sum_{k=j}^{L-i}\left(C^{k-1}_{k-j-1}+C^{k-2}_{k-j}\right)
  Z(L-k,n)= F_1(L,n;i;j),
\end{align}
 where we have used $C^{j-1}_{-1}=0$ (see \eqref{C2}) and
$C^{k-1}_{k-j-1}+C^{k-2}_{k-j}=C^{k-1}_{k-j}$ (see \eqref{C1}).  The case
$j=1$ is easily checked separately.  Similarly one verifies that
 \be
F_2(L,n;i;j+1) +   F_2(L-1,n;i;j-1)= F_2(L,n;i;j)
\ee
 separately for $j\le L-i-1$ and for $j=L-i$.

 \smallskip\noindent
 {\bf Case 2:} We denote the right hand side of \eqref{G} by $G(L,n;i;j)$,
and show that $G$ satisfies the appropriate boundary conditions and
recursion.  One checks immediately that $G(L,n;i;0)=Z(L,n)$ (compare
\eqref{BC1}) and, using \eqref{C2}, that $G(L,n;i;L-n)=1$ (compare
\eqref{BC2/b}).  Finally one shows that, for $1\le j\le L-n-1$,
 \be
G(L,n;i;j+1) + G(L-1,n;i;j-1) = G(L,n;i;j);
 \ee
 the proof is essentially the same as that of the
recursion for $F_1$ in Case~1.  
\end{proof}

\section{The fat shock}\label{sec:fat}

In this section we give a precise definition and analysis of the fat shock
discussed informally in the introduction.  The analysis will be used in the
next section for the determination of local states in the infinite volume
limit.  We define the fat shock microscopically as the region between the
positions $Q_1$ and $Q_n$ of the first and last second class particles in
the system.

 The joint distribution of $Q_1$ and $Q_n$ was obtained in
Remark~\ref{finite}; it is convenient here to write this, for $j,k,l\ge0$,
as
 \begin{align}
 \theta^{\alpha,\beta}_{L,n}(j,k,l)
   &\equiv \mu^{\alpha,\beta}_{L,n}(Q_1=j+1,Q_n=j+k+2)
    \,\delta_{j+k+l,L-2}\nonumber\\
 &=  \frac{Z^{\alpha,1}\,(j,0)Z^{1,1}(k,n-2)\,
      Z^{1,\beta}(l,0)} {Z^{\alpha,\beta}(L,n)}\,\delta_{j+k+l,L-2}.
   \label{thetadef}
 \end{align}
 We can determine the large-$L$ behavior of $\theta$ by replacing the
partition functions in \eqref{thetadef} with their asymptotic forms; these
can be obtained from \cite{dehp} and \cite{arita}, and are summarized in
Appendix~\ref{app:part}.  In some cases it is convenient to further
approximate the distribution of $k$, which represents the fat shock width
on the microscopic scale, by a Gaussian.  (Recall that a macroscopic width
$w=w(\alpha,\beta,\gamma)$ for the fat shock was predicted on heuristic
grounds in Section~1 (see \eqref{wdef}), so we expect that $k\sim wL$ for
large $L$.)  We omit details of the computations and summarize the results
in the next remark; where the notation
$\theta^{\alpha,\beta}_{L,n}(j,k,l)\sim f(\alpha,\beta,\gamma,j,k,l)$
indicates that the ratio of the two quantities approaches 1 as $L\to\infty$
with $n=\floor{\gamma L}$, where $\floor{u}$ is the greatest integer
contained in $u$.

\begin{rem}\label{theta}
  (a) On the boundary of regions II and III ($ \alpha=\beta<\alpha_c $),
 \be\label{thetashock}
 \theta^{\alpha,\alpha}_{L,n}(j,k,l) \sim
   \frac{1}{L(1-w)}\sqrt{\frac{A(\alpha)}{\pi L}}
     \,e^{-A(\alpha)(k-Lw)^2/L}\,\delta_{j+k+l,L-2},
\ee
 where $A(\alpha)=(1-2\alpha)^3/(4\gamma\alpha(1-\alpha))$.  That is, under
$\theta_{L,n}^{\alpha,\alpha}(j,k,l)$, $k$ is approximately Gaussian with
mean $Lw$ and variance of order $L$, $j$ is approximately uniformly
distributed on the range $0\le j\le L-l-2$, and $l=L-2-j-k$.  On the
macroscopic scale, this means that the width of the fat shock is $w$ and
its left endpoint is uniformly distributed on the interval
$[0,1-w]$.

\smallskip\noindent(b)
   In region II ($\alpha<\alpha_c,\alpha<\beta $),
 \be\label{thetaII}
   \theta^{\alpha,\beta}_{L,n}(j,k,l) \sim
  p^{\alpha(1-\alpha),\beta}(l)\,
    \sqrt{\frac{A(\alpha)}{\pi L}}\,e^{-A(\alpha)(k-Lw)^2/L}\,
     \delta_{j+k+l,L-2},
\ee
 where we have introduced the (normalized) probability
distribution
 \be
  p^{u,\beta}(l)=\frac{\beta(1+\sqrt{1-4u})-2u}{2\beta}\,
       u^lZ^{\beta,1}(l,0),\quad l=0,1,\ldots,
 \ee
 defined for $u<\beta(1-\beta)$ if $\beta\le1/2$, $u\le1/4$ otherwise.
$p^{u,\beta}(l)$ decreases exponentially for large $l$ unless $u=1/4$, when
the decrease is as $l^{-3/2}$ (see \eqref{z1alpha}); $p$ is normalized by
\eqref{genfct}.  Thus on the microscopic scale $l$ is typically of order 1
and the shock width $k$ is distributed as in (a).  On the macroscopic scale
the fat shock has width $w$ and is pinned to the right
end of the system.  The analysis in region III is similar.

\smallskip\noindent(c)
   On the boundary of regions I and II ($\alpha_c = \alpha < \beta$),
 \be\label{thetaI/II}
   \theta^{\alpha,\beta}_{L,n}(j,k,l) \sim
  p^{\alpha(1-\alpha),\beta}(l)\,
  2 \sqrt{\frac{A(\alpha)}{\pi L}}\,e^{-A(\alpha)(k-L)^2/L}
  \,\delta_{j+k+l,L-2}.
\ee
 This is as in (b) except that here $w=1$ and as a result $k$ is
distributed as a Gaussian random variable conditioned to  have value at
most equal to its mean, and there is a corresponding factor of $2$ in the
normalization.  The analysis on the I/III boundary is similar.

\smallskip\noindent(d)
   At the triple point ($\alpha_c = \alpha = \beta$),
 \be\label{thetatrip}
   \theta^{\alpha,\beta}_{L,n}(j,k,l) \sim
 \frac{2\gamma^2}{L(1-\gamma)^2}\,e^{-A(\alpha)(k-L)^2/L}\,\delta_{j+k+l,L-2}.
\ee
  The distribution of $k$ is as in (c) but here $j$ and $l$ are free,
  subject only to the constraint $j+l=L-2-k$.

\smallskip\noindent(e)
 In region I ($\alpha_c < \alpha,\beta$),
 \be\label{thetaI}
   \theta^{\alpha,\beta}_{L,n}(j,k,l) \sim
  p^{(1-\gamma^2)/4,\alpha}(j)\,
  p^{(1-\gamma^2)/4,\beta}(l)\,\delta_{j+k+l,L-2};
\ee
 $j$ and $l$ are both of order 1 (microscopically) and $k$ is determined by
the delta function constraint.

  \smallskip\noindent
 Note that the results of Remark~\ref{theta} confirm the picture of the fat
shock behavior sketched in Section~1.
\end{rem}

\section{Local states in the infinite volume limit in the bulk} 
  \label{sec:localbulk}

 In this section we discuss a question inspired by the treatment of the one
component system by Liggett \cite{ligg2}: is there a {\it local state} at
position $x$ of the system (in the infinite volume limit), and if so what
is it?  To formulate a precise question we consider a limit in which $n$
and $i$ increase with $L$ in such a way that $i\to\infty$, $L-i\to\infty$,
$i/L\to x\in[0,1]$, and $n/L\to\gamma\in(0,1)$.
In this setting we ask about the existence and nature of the weak limit
$\lim_{L\to\infty} T^{-i}\mu_{L,n}^{\alpha,\beta}$, where $T$ is
translation by one lattice site and so the operator $T^{-i}$ carries site
$i$ to the origin;   equivalently, we consider the sites of our open system
to run from $1-i$ to $L-i$ and look at the probabilities of configurations
in the interval from $-K$ to $K$, take $L$, $i$, $L-i$, and $n$ to infinity
as above, and then make $K$
arbitrary.  The limit (if it exists) is a measure on the configuration
space $Y=\{0,1,2\}^{\bZ}$; we call it a {\it local state in the bulk} since
(in the $L\to\infty$ limit) it describes a situation infinitely far from
each boundary; the {\it local state at the boundary} will be discussed in
Section~\ref{sec:localbdries}.

It will suffice to consider a special class of these limiting procedures;
specifically, we will always take
 \be\label{choices}
 n = n_L=\floor{\gamma L}\quad\text{and}\quad i = i_L=\floor{xL} + c\sqrt L;
 \ee
 we must assume that $c>0$ if $x=0$ and $c<0$ if $x=1$. We then define
 \be
 \mu_{x,c}\equiv \lim_{L\to\infty} T^{-i_L}\mu_{L,n_L}^{\alpha,\beta}.
    \label{limit}
\ee
 The limit in \eqref{limit} certainly exists along subsequences, by the
compactness of the set of measures on $Y$.  To simplify notation we will
ignore the necessity of passing to subsequences; since the limiting measure
will be found to be unique, the limit of the sequence itself must also
exist.  For most values of the parameters the limit will in fact be
independent of the choice of $c$, but this is not true when $x=x_0$ in
region II or on the I/II boundary, or $x=x_1$ in region III or on the I/III
boundary. 

 We first consider the currents and densities in the state $\mu_{x,c}$.  The
 currents in the finite system, and hence also in the limit, are
 independent of the site:
 \be
 \E{\mu_{x,c}}{\eta_1(j)(1-\eta_1(j+1))}
   =\E{\mu_{x,c}}{(1-\eta_0(j-1))\eta_0(j)} =  J_1 \label{current}
\ee
 for any $j$, with $J_1$ given in \eqref{Jdef}.  The limiting densities
$\rho_a(x,c)$ for $a=0,1,2$ are defined by
 \be
  \rho_a(x,c)
  =\lim_{L\to\infty} \E{\mu_{L,n_L}^{\alpha,\beta}}{\eta_a(i_L)}
    =\E{\mu_{x,c}}{\eta_a(0)},
  \label{limit3}
\ee
 with the last equality expressing the fact that $i_L$ corresponds to the
origin in $\mu_{x,c}$.  It is easy to check, from the asymptotic
computations of \cite{arita}, that the limit in \eqref{limit3} would be
unchanged if $i_L$ were replaced by $i_L+j$ for any fixed $j$, which
implies that $\E{\mu_{x,c}}{\eta_a(j)}=\rho_a(x,c)$ for any $j$, i.e., the
densities under $\mu_{x,c}$ are translation invariant.  Equation
\eqref{limit3} may be viewed as a refined version of \eqref{rhodef}, in
which the ambiguity in the $L\to\infty$ limit there has been removed.

   Noting that in the $L\to\infty$ limit the generator of the dynamics in
the neighborhood of $i_L$ does not involve any boundary terms or any
constraints on the densities of the three species beyond
$\sum_a\eta_a(j)=1$, one verifies easily \cite{ligg2} that $\mu_{x,c}$ must be
invariant for the infinite-volume two species TASEP dynamics.  It then 
follows that $\mu_{x,c}$ must be a convex combination of the extremal invariant
measures for the infinite volume two species TASEP.  These measures have
been classified in \cite{speer}: there is (i)~a family of translation
invariant measures $\nu^{\lambda_0,\lambda_1}$, defined for
$\lambda_0,\lambda_1\ge0$, $\lambda_0+\lambda_1\le1$, in which holes, first
class particles, and second class particles have densities $\lambda_0$,
$\lambda_1$, and $1-\lambda_0-\lambda_1$, respectively, and (ii)~a family
of non-translation-invariant ``blocking'' measures $\hat\nu^{m,n}$, where
$m,n\in\bZ\cup\{-\infty,\infty\}$, $m\le n$, and $m,n$ are not both
infinite: $\hat\nu^{m,n}$ is a unit point mass on the configuration
$\tau^{m,n}$ given by
 \be
  \tau^{m,n}_i
  = \begin{cases}0,& \text{if $i<m$,}\\ 2,&\text{if $m\le i<n$,}\\
     1,& \text{if $n\le i$.}\end{cases}
 \ee
 However, the translation invariance of the densities implies that none of
the blocking measures can be present in the superposition giving
$\mu_{x,c}$.  

Thus there exists a probability measure
$\kappa_{x,c}(d\lambda_0,d\lambda_1)$ (which depends also on $\alpha$,
$\beta$, and $\gamma$) that specifies the weights of the different
translation invariant extremal measures which enter into the superposition:
 \be
 \mu_{x,c}= \int\nolimits_{\lambda_0,\lambda_1\ge0,\ \lambda_0+\lambda_1\le1}
     \kappa_{x,c}(d\lambda_0,d\lambda_1)\,\nu^{\lambda_0,\lambda_1}.
    \label{sup}
\ee
We will see that: (i)~for most values of the parameters, $\kappa_{x,c}$ is a
point mass, so that $\mu_{x,c}$ is one of the measures
$\nu^{\lambda_0,\lambda_1}$, (ii)~in some cases, in which $x$ may lie to
the left of, within, or to the right of the fat shock discussed in the
introduction, $\mu_{x,c}$ is a superposition of the two or three measures
corresponding to these possibilities, and (iii)~no more complicated
superposition can occur.  Note that, as a consequence, the current $J_1$ is
the same for all elements of the superposition (and the same holds for
$J_0$ and for $J_2=0$). Here is a first result in this direction.
  
 \begin{thm}\label{main} If $\mu_{x,c}$ is defined by \eqref{limit} and the
current and densities at $x$ are related by
$J_1=\rho_0(x,c)(1-\rho_0(x,c))=\rho_1(x,c)(1-\rho_1(x,c))$, then
$\mu_{x,c}=\nu^{\rho_0(x,c),\rho_1(x,c)}$.\end{thm}

 The condition in the theorem that
$\rho_0(x,c)(1-\rho_0(x,c))=\rho_1(x,c)(1-\rho_1(x,c))$ corresponds to the
zero current of second class particles and leads to the alternatives of
\eqref{alt}.  We see from Table~\ref{table:dens} that this theorem
determines $\mu_{x,c}$ completely for most but not all values of $\alpha$,
$\beta$, $\gamma$, $x$, and $c$ and that the results are consistent with
the intuitive picture sketched in the introduction. In summary:

 \begin{rem}\label{where} It follows from Theorem~\ref{main}  that:

 \smallskip\noindent
  (a)~In region I the local state $\mu_{x,c}$ is $\nu^{\alpha_c,\alpha_c}$;
 in particular, it is independent of $x$ and $c$.  

 \smallskip\noindent
 (b)~In region II the local state $\mu_{x,c}$ is $\nu^{1-\alpha,\alpha}$
for $x<x_0$ and $\nu^{\alpha,\alpha}$ for $x>x_0$, i.e., respectively
outside and inside the fat shock. Region III is similar: the local state is
$\nu^{\beta,\beta}$ for $x<x_1$ and $\nu^{\beta,1-\beta}$ for $x>x_1$.

 \smallskip\noindent
 Other cases are not determined by the theorem:

 \smallskip\noindent
 (c)~The local state is not determined by Theorem~\ref{main} in the
interior of the regions where either type~2 or type~0 particles have a
linear profile, that is, on the II/III  boundary (where the fat shock is
not pinned to one or the other end of the system) with $0<x<x_0$ or
$1>x>x_1$.  See Figure~\ref{fig:profs}(c,d).
 
 \smallskip\noindent
 (d)~The local state is not determined by Theorem~\ref{main} (i)~at $x=x_0$
in region II and on the I/II boundary, where $x_0=0$; (ii)~at $x=x_1$ in
region III and on the I/III boundary, where $x_1=1$, or (iii)~at $x=x_0=0$
and $x=x_1=1$ at the triple point.  All of these points are edges of the
(pinned) fat shock; see Figure~\ref{fig:profs}(a,b,e).

 \smallskip\noindent
  We will determine $\mu_{x,c}$ in cases (c) and (d) below.
\end{rem}

\begin{proofof}{Theorem~\ref{main}} Using \eqref{sup} together with the
relations $\E{\nu^{\lambda_0,\lambda_1}}{\eta_1(i)}=\lambda_1$,
$\E{\nu^{\lambda_0,\lambda_1}}{\eta_1(i)(1-\eta_1(i+1))}
=\lambda_1(1-\lambda_1)$ (which hold for all $i$ because the marginal of 
$\nu^{\lambda_0,\lambda_1}$ on configurations of first class particles is a product measure), we
find that
 \be
 \rho_1(x,c)=\E{\mu_{x,c}}{\eta_1(i)}
 = \int \lambda_1\,\kappa_{x,c}(d\lambda_0,d\lambda_1)
=\E{\kappa_{x,c}}{\lambda_1},
 \ee
 and 
 \begin{align}
  J_1&=\E{\mu_{x,c}}{\eta_1(i)(1-\eta_1(i+1))}\nonumber\\
  &= \int \lambda_1(1-\lambda_1)\,\kappa_{x,c}(d\lambda_0,d\lambda_1)
    =\E{\kappa_{x,c}}{\lambda_1(1-\lambda_1)}.
\end{align}
 From $J_1=\rho_1(x,c)(1-\rho_1(x,c))$, then, we see that
$\E{\kappa_{x,c}}{\lambda_1^2}=\E{\kappa_{x,c}}{\lambda_1}^2$, so that
$\lambda_1=\E{\kappa_{x,c}}{\lambda_1}=\rho_1(x,c)$ almost surely with
respect to $\kappa_{x,c}$.  Similarly, $\lambda_0=\rho_0(x,c)$ almost
surely with respect to $\kappa_{x,c}$, so that
$\mu_{x,c}=\nu^{\rho_0(x,c),\rho_1(x,c)}$.  \end{proofof}

We now turn to the determination of the local measure $\mu_{x,c}$ at a
point $x$ where the densities are varying linearly or are
discontinuous---case (c) or (d) of Remark~\ref{where}.  Recall that in
Section~\ref{sec:fat} we have determined the probability
$\theta^{\alpha,\beta}_{L,n}(j,k,l)$ that $Q_1=j+1$ and $Q_n=j+k+2$, where
$j+k+l=L-2$ and $Q_1$ and $Q_n$ are the position of the first and last
second class particles.  Now let $\mu_{L,n,j,k,l}^{\alpha,\beta}$ denote
the measure $\mu_{L,n}^{\alpha,\beta}$ conditioned on $Q_1=j+1$,
$Q_n=j+k+2$. The key observation we will use follows from
Remark~\ref{finite}, specifically, from \eqref{factortau} or a simple
generalization thereof: if $G$ is a function on $Y_{L,n}$ which depends on
the $\tau_i$ only for $m_0\le i\le m_1$, then
 \begin{equation}\label{key}
\E{\mu_{L,n,j,k,l}^{\alpha,\beta}}{G}=\begin{cases}
     \E{\mu_{j,0}^{\alpha,1}}{G},& \text{if $m_1\le j$},\\
     \E{\mu_{k,n-2}^{1,1}}{T^{-j-1}G},
            & \text{if $j+2\le m_0,\; m_1\le i+k+1$},\\
     \E{\mu_{l,0}^{1,\beta}}{T^{-j-k-2}G},& \text{if $j+k+3\le m_0$}.
\end{cases} 
 \end{equation}

 Now let $(r_L)$ be a sequence of integers with $r_L\to\infty$ and
$r_L/\sqrt{L}\to0$.  For any function $F$ on $Y$ depending only on finitely
many spins we may write
 \begin{align}\label{decomp}
 \E{\mu_{x,c}}{F}  &=   \lim_{L\to\infty}
   \sum_{j+k+l=L-2} \theta_{L,n_L}^{\alpha,\beta}(j,k,l)
    \E{\mu_{L,n_L,j,k,l}^{\alpha,\beta}}{T^{i_L}F} \nonumber\\
 &=   \lim_{L\to\infty} \bigl[\Xi^{(1)}_L\E{\mu^{(1)}_L}{F}
    +\Xi^{(2)}_L\E{\mu^{(2)}_L}{F}+\Xi^{(3)}_L\E{\mu^{(3)}_L}{F}
  + {\rm remainder}\bigr].
\end{align}
 Here $\mu_L^{(p)}$ is for $p=1,2,3$ the probability measure defined by 
 \be\label{mup}
\mu^{(p)}_L=\Xi^{(p)}_L\raise2pt\hbox{${}^{-1}$}\sum\nolimits^{(p)}_{j,k,l}
  \theta_{L,n_L}^{\alpha,\beta}(j,k,l)
   T^{-i_L}\mu_{L,n_L,j,k,l}^{\alpha,\beta},
 \ee
 where  $\sum^{(1)}_{j,k,l}$ ranges over values satisfying $j>i_L+r_L$,
$\sum^{(2)}_{j,k,l}$ over $j<i_L-r_L$ and $k>i_L+r_L$, $\sum^{(3)}_{j,k,l}$
over $k<i_L-r_L$, and 
 \be
\Xi^{(p)}_L  = \sum\nolimits^{(p)}_{j,k,l}
  \theta_{L,n_L}^{\alpha,\beta}(j,k,l)
  =\begin{cases}\label{Xidef}
      \mu^{\alpha,\beta}_{n,L}(Q_1>i_L+r_L),& \text{if $p=1$},\\
      \mu^{\alpha,\beta}_{n,L}(Q_1<i_L-r_L,\;Q_n>i_L+r_L),& \text{if $p=2$},\\
      \mu^{\alpha,\beta}_{n,L}(Q_n<I_L-r_L),& \text{if $p=3$}.\end{cases}
 \ee
  The remainder in \eqref{mup} contains those terms from the full sum over
$i$ and $k$ which are omitted from $\sum^{(1)}$, $\sum^{(2)}$, and
$\sum^{(3)}$.  We have suppressed  the dependence of these various entities
  on $\alpha$, $\beta$, $\gamma$, $x$, and $c$.

We now take the $L\to\infty$ limit in \eqref{decomp}.  It follows from
Remark~\ref{theta} and the fact that $r_L$ grows more slowly than $\sqrt{L}$
that the remainder vanishes in this limit.  The $\Xi^{(p)}_L$ are expressed
as probabilities in \eqref{Xidef} and their limiting values
$\Xi_{x,c}^{(p)}=\lim_{L\to\infty}\Xi^{(p)}_L$ may be determined from
Remark~\ref{theta}; these limits will be discussed on a case by case basis
below.

 Finally, to study $\lim_{L\to\infty}\mu_L^{(p)}$ we observe that for
sufficiently large $L$ (if $F$ depends on $\tau_i$ only for $|i|\le m$ then
$r_L>m$ suffices) we have by \eqref{key} that
 \be\label{mup2}
\E{\mu_L^{(p)}}{F}=\begin{cases}
  \Xi^{(1)}_L\raise2pt\hbox{${}^{-1}$}\sum\nolimits^{(1)}_{j,k,l}
 \theta_{L,n_L}^{\alpha,\beta}(j,k,l)\E{\mu_{j,0}^{\alpha,1}}{T^{i_L}F},
        & \text{if $p=1$},\\\noalign{\medskip}
  \Xi^{(2)}_L\raise2pt\hbox{${}^{-1}$}\sum\nolimits^{(2)}_{j,k,l}
  \theta_{L,n_L}^{\alpha,\beta}(j,k,l)\E{\mu_{k,n-2}^{1,1}}{T^{i_L-j-1}F},
        & \text{if $p=2$},\\\noalign{\medskip}
  \Xi^{(3)}_L\raise2pt\hbox{${}^{-1}$}\sum\nolimits^{(3)}_{j,k,l}
  \theta_{L,n_L}^{\alpha,\beta}(j,k,l)\E{\mu_{l,0}^{1,\beta}}{T^{i_L-j-k-2}F},
         & \text{if $p=3$}.
 \end{cases} 
 \ee
 The limits $\lim_{L\to\infty}\mu_L^{(p)}$ for $p=1,2,3$ are all treated
similarly; let us discuss the case $p=2$ in detail.  Equation \eqref{mup2}
displays $\mu_L^{(2)}$ as a convex combination of the probability measures
$T^{-(i_L-j-1)}\mu_{k,n-2}^{1,1}$.  Each of these measures, for large $L$,
is by Remark~\ref{where}(a) approximately equal to
$\nu^{\alpha\wedge\beta,\alpha\wedge\beta}$ (recall that
$\alpha\wedge\beta=\min\{\alpha,\beta\}$), since the critical value
$\alpha_c^*$ of $\alpha$ for a system with
$k\simeq wL=\gamma L/(1-2\alpha\wedge\beta)$ sites and $n\simeq \gamma L$
second class particles---and thus an effective value
$\gamma^*=n/k=1-2\alpha\wedge\beta$ of $\gamma$---is
$(1-\gamma^*)/2=\alpha\wedge\beta$.  The same should be true of
$\mu_L^{(2)}$.  The corresponding evaluations for $p=1,3$ come from the
results of \cite{ligg2} for the local measures in the one species open
system.  We conclude that
 \be\label{limmu}
\lim_{L\to\infty}\mu_L^{(p)}=\begin{cases}
   \nu^{1-\alpha\wedge(1/2),\alpha\wedge(1/2)},& \text{if $p=1$},\\ 
   \nu^{\alpha\wedge\beta,\alpha\wedge\beta},& \text{if $p=2$},\\ 
   \nu^{\beta\wedge(1/2),1-\beta\wedge(1/2)},& \text{if $p=3$},\end{cases}
 \ee
and so
 \be\label{localsup}
 \mu_{x,c}=\Xi_{x,c}^{(1)}\nu^{1-\alpha\wedge(1/2),\alpha\wedge(1/2)}+
 \Xi_{x,c}^{(2)}\nu^{\alpha\wedge\beta,\alpha\wedge\beta}+
 \Xi_{x,c}^{(3)}\nu^{\beta\wedge(1/2),1-\beta\wedge(1/2)}.
\ee

\begin{rem}\label{rem:proof}
The heuristic argument for \eqref{limmu} given above could be
made precise by justifying the implicit exchange of limits;  we sketch
instead an alternate proof, again for $p=2$.  We know that the limiting
current for the measures $\mu^{1,1}_{k,n-2}$, as $L$ and hence $k\simeq wL$
goes to infinity, is $\alpha(1-\alpha)$, and the limiting densities of
holes, particles, and second class particles are $\alpha$, $\alpha$, and
$1-2\alpha$, respectively.  One can in fact show further that these limits
are obtained with  error which goes to zero uniformly at sites $i$ satisfying
$r_L\le i\le k-r_L$; from this, it follows that the measures
$\mu_L^{(2)}$ have the same limiting current and densities.  Then an
argument as in the proof of Theorem~\ref{main} establishes \eqref{limmu}.
\end{rem}

To complete our discussion of the local states in the bulk we must
supplement \eqref{localsup} with a determination of the weights
$\Xi_{x,c}^{(p)}\equiv \lim_{L\to\infty} \Xi^{(p)}_L$ for cases (c) and (d)
of Remark~\ref{where}.  The cases in the next remark are parallel to those
of Remark~\ref{theta}.

\begin{rem}\label{Xis} (a) On the boundary of regions II and III (the shock
line, case (c) of Remark~\ref{where}) we find from Remark~\ref{theta}(a)
and \eqref{mup2} that
 \be\label{xiinfa}
 \Xi_{x,c}^{(p)}= \frac1{1-w}\times
  \begin{cases} (1-w-x)_+,& \text{if $p=1$,}\\
  1-w-(1-w-x)_+-(x-w)_+,&\text{if $p=2$,}\\
  (x-w)_+,& \text{if $p=3$.}
\end{cases}
\ee
 Here $u_+=u$ if $u\ge0$ and $u_+=0$ if $u<0$.  Note that these
coefficients, and hence the local measure $\mu_{x,c}$ given by
\eqref{localsup}, are independent of $c$.

 \smallskip\noindent
 (b) In region II ($\alpha<\alpha_c,\alpha<\beta $), at the fixed shock
$x_0$, the $\Xi_{x,c}^{(p)}$ do depend on $c$: 
 \be\label{xiinfb}
 \Xi_{x_0,c}^{(1)}= 1-\Phi\left(c\sqrt{A(\alpha)}\right),\qquad
 \Xi_{x_0,c}^{(2)}= \Phi\left(c\sqrt{A(\alpha)}\right),\qquad
 \Xi_{x_0,c}^{(3)}= 0.
\ee
 Here $\Phi$ is the error function defined by 
 \be\label{erf}
 \Phi(t)=\frac{1}{\sqrt{2\pi}}\int_{-\infty}^te^{-\tau^2/2}\,d\tau.
  \ee
 The analysis in region III is similar.

\smallskip\noindent
(c) On the boundary of regions I and II ($\alpha_c = \alpha < \beta$),
where $x_0=0$, we find that for  $c>0$,
 \be\label{xiinfc}
 \Xi_{0,c}^{(1)}= 2-2\Phi\left(c\sqrt{A(\alpha)}\right),\qquad
 \Xi_{0,c}^{(2)}= 2\Phi\left(c\sqrt{A(\alpha)}\right)-1,\qquad
 \Xi_{0,c}^{(3)}= 0.
\ee
 The analysis on the I/III boundary is similar: for $c<0$, 
 \be\label{xiinfcp}
 \Xi_{1,c}^{(1)}= 0,\qquad
 \Xi_{1,c}^{(2)}= 1-2\Phi\left(c\sqrt{A(\alpha)}\right),\qquad
 \Xi_{1,c}^{(3)}= 2\Phi\left(c\sqrt{A(\alpha)}\right).
\ee

\smallskip\noindent
(d) At the triple point ($\alpha_c = \alpha = \beta$), where $x_0=0$ and
$x_1=1$, \eqref{xiinfc} and \eqref{xiinfcp} again hold (with $c>0$ and
$c<0$ respectively).
\end{rem}

We finally note that the results of this section yield density profiles as
well as the finite volume corrections to these profiles at the fixed shocks
(see Remark~\ref{rem:nonunique}(a)), since $\rho_a(x,c)$ may be calculated
from \eqref{limit3} and \eqref{localsup}.  For example, on the II/III border
(shock line) we find in this way that
 \be\label{rholin}
 \rho_0(x)
  = \Xi_{x,c}^{(1)}(1-\alpha)+(\Xi_{x,c}^{(2)}+\Xi_{x,c}^{(3)})\alpha,\quad
 \rho_1(x)= (\Xi_{x,c}^{(1)}+\Xi_{x,c}^{(2)})\alpha+\Xi_{x,c}^{(3)}(1-\alpha),
\ee
with the weights $\Xi^{(p)}$ given by \eqref{xiinfa}; it is easy to see
that \eqref{rholin} reproduces \eqref{linear}. Here we have used the
notation $\rho_a(x)$ of \eqref{rhodef}, rather than writing $\rho_a(x,c)$ as
in \eqref{limit3}, since the densities do not depend on $c$.  In region II
we have, from \eqref{localsup} and \eqref{xiinfa}, that
 \be \label{rhoshock}
 \rho_0(x_0,c) = 1-\alpha-\Phi\left(c\sqrt{A(\alpha)}\right)(1-2\alpha). 
 \ee
 Similar results hold in region III at the point $x_1$. 

\section{Local states in the infinite volume limit at the boundaries}
\label{sec:localbdries}

In this section we study limiting measures
$\lim_{L\to\infty} T^{-i_L}\mu_{L,n_L}^{\alpha,\beta}$ as in \eqref{limit},
still taking $n_L=\floor{\gamma L}$ but now assuming that either $i_L$ or
$L-i_L$ is fixed.  Without loss of generality we can assume that $i_L=1$ or
$i_L=L$  (the measure as seen from site $j$ or site $L-j$ can be recovered
from these limits) and thus define
 \be
 \mu_{\rm left}\equiv \lim_{L\to\infty} 
  T^{-1}\mu_{L,\floor{\gamma L}}^{\alpha,\beta},\qquad
 \mu_{\rm right}\equiv \lim_{L\to\infty} T^{-L}
  \mu_{L,\floor{\gamma L}}^{\alpha,\beta}.
    \label{limitbd}
\ee
 Note that $\mu_{\rm left}$ (respectively $\mu_{\rm right}$) does not
coincide with any of the measures $\mu_{0,c}$, $c>0$, (respectively
$\mu_{1,c}$, $c<0$,) studied in Section~\ref{sec:localbulk}.  The densities
under $\mu_{\rm left}$ and $\mu_{\rm right}$ were studied in \cite{arita};
for example, $\E{\mu_{\rm left}}{\eta_a(j)}$ is denoted
$\rho^a_{{\rm left},j}$ in \cite{arita}.

By the particle hole symmetry it suffices to consider $\mu_{\rm left}$,
which is a measure on the semi-infinite configuration space
$\{0,1,2\}^{\{0,1,2,3,\ldots\}}$.  In general we do not have a proof that
the limit defining $\mu_{\rm left} $ exists (except along subsequences),
although we expect this to be true; see also the comment below
Theorem~\ref{bern}.  The next result, however, gives a somewhat surprising
property which any (subsequence) limit must satisfy; to simplify notation,
we will speak as if the limit itself exists.

\begin{thm}\label{bern} The distribution of first class particles under the
measure $\mu_{\rm left}$ is Bernoulli with a constant density $\rho$, where
$\rho$ is given by
\be \label{dens}
\rho = \begin{cases} 
\alpha_c, & \text{in region I,} \\
\alpha, & \text{in region II,} \\
\beta, & \text{in region III.} 
\end{cases}
\ee
\end{thm}

\begin{proof} By Theorem~\ref{thm:exch}, we know that the (marginal)
distribution of the variables $\eta_1(i)$ under $\mu_{\rm left}$ is
exchangeable, so that by de Finetti's theorem \cite{feller} this
distribution is a superposition of Bernoulli distributions.  From
\cite{arita} we know that for any $i\ge0$,
$\rho\equiv\E{\mu_{\rm
left}}{\eta_1(i)}=\lim_{L\to\infty}\E{\mu^{\alpha,\beta}_{L,\floor{\gamma
L}}}{\eta_1(i)}$
is given by \eqref{dens} and that
$\lim_{L\to\infty}\E{\mu^{\alpha,\beta}_{L,\floor{\gamma
L}}}{\eta_1(i)(1-\eta_1(i+1)}=J_1$
(see \eqref{Jdef}). In each region of the phase plane these limits satisfy
the relation $J=\rho(1-\rho)$. Then from an argument as in the proof of
Theorem~\ref{main} it follows that the $\eta_1(i)$ distribution is the
product measure $\nu^\rho$, where $\rho$ is given by \eqref{dens}.
\end{proof}

Note that in region II the density of second class particles any finite
distance from the left boundary goes to zero as $L\to\infty$ \cite{arita},
so that knowing that the distribution of particles is Bernoulli completely
determines any limiting measure to be a Bernoulli measure on particles and
holes only.  It follows that the limiting measure exists without passing to
subsequences.

We discuss briefly one aspect of the measure $\mu_{\rm left}$ in the limit
$\gamma\to0$ (note that we are taking this limit {\it after} the
$L\to\infty$ limit).  Consider first region~I; from Remark~\ref{theta}(e)
we see that the position $Q_1$ of the first second class particle in the
system is distributed according to $p^{1/4,\alpha}(q_1)$; this is a
normalizable distribution which decreases as $q_1^{-3/2}$ for large $q_1$,
so that there remains a second class particle in the system, but
$\E{\mu_{\rm left}}{Q_1}=\infty$. In fact more is true; by a calculation
similar to that of Remark~\ref{theta} one can show that all $Q_j-Q_{j-1}$,
$j=2,3,,\ldots$, have this same distribution (see also the discussion of the
pressure ensemble in Section~\ref{sec:equil}) so that there remain an
infinite number of second class particles in the system under
$\mu_{\rm left}$.  The same is true in Region III, but there by
Remark~\ref{theta}(b) $Q_1$ is distributed according
to $p^{\beta(1-\beta),\alpha}$, so that $\E{\mu_{\rm left}}{Q_1}<\infty$;
the distribution of the $Q_j-Q_{j-1}$, $j=2,3,,\ldots$, is the same as in
Region~I.

\begin{rem} \label{djlsbern} One may compare Theorem~\ref{bern} with result
in \cite{djls} for the infinite volume limit of a two-component TASEP
system on a ring: that the distribution of first class particles to the
right of a second class particle, and that of holes to the left of such a
particle, is Bernoulli.  The two results are closely related, because if we
set $\alpha=\beta=1$ in the open system then the matrix element
$\langle W_1|X_{\tau_1}\cdots X_{\tau_L}|V_1\rangle$ giving the weight of
the configuration $\tau_1,\ldots,\tau_L$ is \cite{djls} exactly the weight
of the configuration $2,\tau_1,\ldots,\tau_L$ on a ring.  Because the
numbers of first class particles and of holes on the ring is fixed, and
these numbers fluctuate in the open system, this does not establish an
exact equivalence of the $\alpha=\beta=1$ case of Theorem~\ref{bern} to the
result of \cite{djls}; nevertheless, it is clear that the former is in some
sense a generalization of the latter to values of $\alpha$ and $\beta$
other than 1.  But the result of \cite{djls} is in another sense more
general than Theorem~\ref{bern}, since the infinite volume limit of the
open system has zero current of second class particles, but this is not
true for the system of \cite{djls}.  \end{rem}

\section{The pressure ensemble for second class particles} \label{sec:equil}

We here consider the steady state distribution of the second
class particles only, so that  one may think of identifying the first
class particles and holes as a new type of hole.  For $d$ a positive
integer we define
 \be\label{phidef}
\phi_\alpha(d)=-\log (4^{-d} \Z^{\alpha,1}(d-1,0))
  = -\log (4^{-d}\EE{W_1|(X_0+X_1)^{d-1}|V_1})
\ee
It follows from the $(\alpha,\beta)$ symmetry of $Z^{\alpha,\beta} (L,n)$
that $\phi_\beta(d)$ is also equal to $-\log(4^{-d}Z^{1,\beta}(d-1,0))$.
Using \eqref{factorZ} we find that the probability that the $n$ second class
particles in the systems are located at sites $q_1<q_2<\cdots<q_n$ is
 \begin{multline}\label{equil}
 \mu_{L,n}^{\alpha,\beta}(Q_1=q_1,\ldots,Q_n=q_n)\\
 = (4^{-L}\Z_{\alpha,\beta}(L,n))^{-1}
  e^{-\phi_\alpha(q_1)
   -\sum_{i=2}^n\phi(q_i-q_{i-1})-\phi_\beta(L-q_n)},
 \end{multline}
% \ee
 %
 where we have denoted $\phi_1(d)$ by $\phi(d)$.  The motivation for the
factors $4^{-d}$ in \eqref{phidef} will be discussed below; with this
normalization $\phi(d)\sim (3/2)\log d$ for $d\to\infty$ \cite{djls,ffk}.

We note that \eqref{equil} has precisely the form of the canonical
distribution for a system in a domain of length $L$ with particles
interacting with their nearest neighbor only via a pair potential
$\phi(d)$. (Such an interaction is rather unphysical; one may think of any
intervening particle as screening the interaction of particles that it
separates.). There is also a potential $\phi_\alpha(d)$ ($\phi_\beta(d))$
representing the interaction of the first (last) particle with the left
(right) boundary.

 The TASEP dynamics for the full system gives rise in a natural way to a
dynamics on the system of the second class particles which satisfies
detailed balance with respect to this equilibrium measure.  In the state in
which the second class particles are at $q_1,\ldots,q_n$ the $i^{\rm th}$
second class particle moves to site $q_i+1$ at rate 1 whenever that site is
empty (in the original sense), an event which by a simple generalization of
\eqref{factortau} occurs in the NESS with probability
 \be\label{rateright}
\frac{\langle W_1| X_0 (X_0+X_1)^{q_{i+1}-q_i-2}|V_1 \rangle}
   {Z^{1,1}(q_{i+1}-q_i-1,0)}
 =\frac{e^{-\phi(q_{i+1}-q_i-1)}}{e^{-\phi(q_{i+1}-q_i)}},
 \qquad\hbox{if $i<n$,}
 \ee
 and with probability
 \be
\frac{\langle W_1| X_0 (X_0+X_1)^{L-q_i-1}|V_\beta \rangle}
   {Z^{1,\beta}(L-q_i,0)}
 =\frac{e^{-\chi_\beta(L-q_i-1)}}{e^{-\chi_\beta(L-q_i)}},
 \qquad\hbox{if $i=n$}.
 \ee
 One finds similarly that the probability that the site $q_i-1$ is occupied
by a first class particle is
 \begin{equation}\label{rateleft}
\frac{e^{-\phi(q_{i}-q_{i-1}-1)}}{e^{-\phi(q_{i}-q_{i-1})}},\quad
  \hbox{if $i>1$},\qquad
\frac{e^{-\psi_\alpha(q_{1}-1)}}{e^{-\psi_\alpha(q_{i})}},\quad
  \hbox{if $i=1$}.
 \end{equation}
 The dynamics in which $q_i\to q_i+1$ when $q_{i+1}-q_i\ge2$, with rate
given by \eqref{rateright}, and $q_i\to q_i-1$ when $q_i-q_{i-1}\ge2$,
with rate given by \eqref{rateleft}, is easily seen to satisfy detailed
balance with respect to the  measure \eqref{equil}.

To obtain the properties of the system described by \eqref{equil} in the
thermodynamic limit, $L \to \infty, n/L \to \gamma$, it is most convenient
to consider the {\it pressure} or {\it isobaric ensemble}
$\pi_{p,n}^{\alpha,\beta}$ \cite{hill,percus}, in which instead of keeping
the volume $L$ of the system fixed we imagine that the right wall is in
contact with a reservoir of pressure $p$.  The value of $p$ can be chosen
so as to make the average volume equal to $L$, as discussed
below. More precisely, we let the position of the right boundary, which we
denote $q_{n+1}$, fluctuate, and add a term involving the pressure $p$ to
the measure. This yields the probability distribution in the pressure
ensemble:
\be
 \begin{split}\label{pidef}
 \pi_{p,n}^{\alpha,\beta}(q_1,\ldots,q_n,q_{n+1}) &\\ 
 &\hskip-87pt = \cZ^{\alpha,\beta}(p,n)^{-1}
\exp\left(-\phi_\alpha(q_1)
   -\sum_{i=2}^n\phi(q_i-q_{i-1})-\phi_\beta(L-q_n)-pq_{n+1}\right).
 \end{split}
\ee
 The partition function has the form
$\cZ^{\alpha,\beta}(p,n)=\cZ_1(\alpha,p)\cZ_2(p)^n\cZ_1(\beta,p)$,
where $\cZ_1$ and $\cZ_2$ are readily found, for $z=\sqrt{1-e^{-p}}$
satisfying
 \be\label{zbds}
  1\ge z\ge\max\{0,1-2\alpha, 1-2\beta\},
 \ee
 to be given by
 \begin{equation} \label{partfn}
\cZ_1(\alpha,p)=\frac{\alpha(1-z)}{z+2\alpha-1}, \qquad
   \cZ_2(p) = \frac{1-z}{1+z}.
 \end{equation}
 Thus \eqref{pidef} becomes
\be\label{pi2}
 \begin{split}
 \pi_{p,n}^{\alpha,\beta}(q_1,\ldots,q_n,q_{n+1}) 
  &= \cZ_1(\alpha,p)^{-1}e^{-\phi_\alpha(q_1)-p q_1}\\
  &\hskip-80pt \times \left[\prod_{i=2}^n 
   \cZ_2(p)^{-1} e^{-\phi_\alpha(q_i-q_{i-1})-p (q_i-q_{i-1})} \right] 
   \cZ_1(\beta,p)^{-1} e^{-\phi_\beta(q_{n+1}-q_n)-p(q_{n+1}-q_n)}.
 \end{split}
\ee
The convenient factorization property of the probability
$\pi_{p,n}^{\alpha,\beta}(q_1,\ldots,q_n,q_{n+1})$ displayed in \eqref{pi2},
which implies that the variables $q_1$ and $q_j-q_{j-1}$, $j=2,\ldots,n+1$,
are independent, has prompted the use of the pressure ensemble for
equilibrium systems, without any reference to dynamics.  The requirement
that particles only interact with their first neighbors is usually imposed
artificially (see, however, \cite{kuh}).  In our model this condition
arises naturally from the dynamics.  Note that $q_n-q_1$, the width of
the fat shock, is thus represented as a sum of independent random
variables; this is consistent with its Gaussian distribution in regions I
and II of the fixed volume ensemble (see Remark~\ref{theta}).

One easily checks that (writing now  $\pi_{p,n}^{\alpha,\beta}=\pi$)
\be\label{exps}
\begin{split}
\E{\pi}{q_1} &= -\frac{d}{dp}\log \cZ_1(\alpha) 
= \frac{\alpha(1+z)}{z(z+2\alpha-1)}, \\
\E{\pi}{q_j-q_{j-1}} 
  &= -\frac{d}{dp}\log \cZ_2 = \frac{1}{z},\qquad j=2,\ldots,n, \\
\E{\pi}{q_{n+1}-q_n} &= -\frac{d}{dp}\log \cZ_1(\beta) 
= \frac{\beta(1+z)}{z(z+2\beta-1)}. 
\end{split}
\ee
 Note that when $z$ approaches its lower bound, which is $0$ if
$\alpha,\beta\ge1/2$ and $\max\{1-2\alpha, 1-2\beta\}$ otherwise, the
average size $\E{\pi}{q_{n+1}}$ of the system goes to infinity for every
$n\ge1$; there is simply not enough pressure to confine the system.  To
agree with standard definitions we have defined the potentials
$\phi_\alpha$ in \eqref{phidef} so the the size of the system in the
absence of boundary terms,
that is, $\E{\pi}{q_n-q_1}$, goes to infinity when $p\to0$ ($z\to0$).

To find the appropriate pressure corresponding to the canonical ensemble
with $L=n/\gamma$ studied above we must set the expected system length
 \be\label{defpress}
  \E{\pi}{q_{n+1}}   
 =  \E{\pi}{q_1}+\sum_{j=2}^n \E{\pi}{q_j-q_{j-1}}+ \E{\pi}{q_{n+1}-q_n}   
  = -\frac{d}{dp}\log \cZ^{\alpha,\beta}(p,n),
   \ee
 equal to $L$ and solve for $p$ (or $z$), subject to the restrictions
\eqref{zbds}.  With \eqref{exps} the equation to be solved becomes
 \be\label{solve}
  \frac{\alpha(1+z)}{z(z+2\alpha-1)}
 +\frac{n}{z} +  \frac{\beta(1+z)}{z(z+2\beta-1)} =\frac n\gamma.
 \ee
 We will discuss the solution of this equation in various regions of the
 phase plane; it is useful to bear in mind that each of the three terms  on
 the left hand side increases as $z$ decreases from 1 to its lower limit 
$\max\{0,1-2\alpha, 1-2\beta\}$.

 Consider first  region~I, where $\alpha,\beta>(1-\gamma)/2$.  Since for
$z=\gamma$ the left hand side of \eqref{solve} is $n/\gamma+O(1)$, where
the $O(1)$ term is positive, the solution must be of the form
$z=\gamma+O(1/n)$.  In the limit $n\to\infty$ we thus have $z=\gamma$ or
$p=-\log(1-\gamma^2)$.  In this case $\E{\pi}{q_1}$ and all
$\E{\pi}{q_j-q_{j-1}}$, $j=2,\ldots,n+1$, are of order unity, so the bulk
of the system has, in the limit $n\to\infty$, the same structure as that
obtained from our NESS when $L\to\infty$ in region~I.

In Region~II, where $\alpha<\beta$ and $\alpha<(1-\gamma)/2$, we have from 
$z>1-2\alpha>1-2\beta$ that  the third term in \eqref{solve} is $O(1)$, and
from $z>1-2\alpha>\gamma$ that the first term must be
$O(n)$, i.e., we must have $z=1-2\alpha+O(1/n)$.  In fact we find easily
that for large $n$, 
 \be
z\simeq 1-2\alpha+\frac{2\alpha(1-\alpha)}{1-2\alpha-\gamma}\,\frac1n.
 \ee
  Now $\E{\pi}{q_1}$ is of order $n$ but $\E{\pi}{q_{n+1}-q_1}$ is still of
order 1; this corresponds to the fat shock being fixed to the right wall,
i.e., to what we see in region II.  The situation in region III is of
course obtained by interchanging $\alpha$ with $\beta$ and left with right.
In the case when $\alpha=\beta<(1-2\gamma)/2$ one sees again that
$z\simeq 1-2\alpha+c/n$ and that $\E{\pi}{q_1}=\E{\pi}{q_{n+1}-q_n}$ are
both of order $n$; this simply means that the average position of the
fat shock is in the middle.

One may of course analyze the pressure ensemble directly, rather than
looking for the correspondence with the open system of fixed length; the
key question is how one allows $p$ or equivalently $z$ to vary with the
number $n$ of (second class) particles.  If $z$ is held fixed (necessarily
in the range \eqref{zbds}) then the behavior of the system corresponds to
that of the open system in region~I.  If $\alpha<1/2$ and $\alpha<\beta$,
and one takes $z=1-2\alpha+c/n$ then the behavior is as in region~II;
similarly if $\beta<1/2$ and $\beta<\alpha$ one obtains region~III behavior
by taking $z=1-2\beta+c/n$, and if $\alpha=\beta<1/2$ such a $z$ value
gives behavior corresponding to the II/III boundary.  A detailed analysis
of the ensemble, for example of the shape of the profiles of the fat shock,
would essentially repeat the analysis in the fixed $L$, i.e., fixed overall
density $\gamma$, ensemble studied earlier, and we will not take
up these questions again here.

  Note that for almost all permissible values of the pressure our system is
in region I; only by fine tuning the pressure to change with $n$ in a range
of order $1/n$ do we get configurations as in regions II or III. This is
reminiscent of what happens when one goes from a fixed magnetization to a
fixed external magnetic field $h$ in the Ising model at low temperatures,
in dimension two or higher.  The whole coexistence region, corresponding to
the average magnetization being smaller than the spontaneous magnetization,
corresponds to the single value $h=0$.

Other choices of $z$ can lead to regimes different from those considered in
the present work.  For example, if we again suppose that
$\alpha=\beta<1/2$, but now take $z$ closer than order $1/n$ to
$1-2\alpha$---to be specific, say $z=1-2\alpha + c/n^2$---then
$\E{\pi}{q_{n+1}}$ is of order $n^2$ and hence the density of second class
particles is zero.

\section{Concluding Remarks} \label{sec:con} 

\noindent 1. As noted already in several places above, the local properties
of our system away from the boundaries approach, as $L \to \infty$, those
of the states of the two species TASEP on the lattice $\bZ$.  Because of
this it will be useful to describe here some known properties of the
(extremal, translation invariant) NESS's of that system, i.e., of the
states $\nu^{\rho_0,\rho_1}$ introduced in Section~\ref{sec:localbulk}.
These states differ from those of other models for which the NESS of the
finite open system can be solved exactly, such as the one species simple
exclusion process and the zero range processes \cite{djls,zrp,bdgjl}, in that
they are not product measures; this is so despite the fact that their
projections (marginals) on the configurations of first class particles
alone, or on the configurations of holes alone, are in fact Bernoulli.  The
states $\nu^{\rho_0,\rho_1}$ may be obtained \cite{djls} as the
$N \to \infty$ limits of states of a two component TASEP on a ring of $N$
sites, with $N_\alpha = \rho_\alpha N$ particles of type $\alpha$,
$\alpha=0,1,2$, where $\rho_2=1-\rho_0-\rho_1$; see also \cite{speer}.

As noted in \cite{djls}, the structure of the states $\nu^{\rho_0,\rho_1}$
is quite intricate, containing several mysterious features which we still
do not understand in any direct intuitive way.  They are not even Gibbs
measures \cite{speer}, even though all the truncated correlation functions
involving a finite number of sites decay exponentially fast.  This decay
follows from the (mysterious) fact that, conditioned on the presence of a
second class particle at site $i$, the measure factorizes: the left and
right sides of $i$ become independent.  For the corresponding property for
the open system studied in this paper see Remark~\ref{finite}.  Another
(mysterious) fact is that if one conditions on
the presence of a second class particle at $i$ then the particles to the
right of $i$, and the holes to the left of $i$, have this Bernoulli
property \cite{djls}. 

Another related property of the states $\nu^{\rho_0,\rho_1}$ is that if one
conditions on there being a first class particle (respectively a hole) at
site $i$ then the measure to the left (respectively right) of $i$ is the
same as if there was no conditioning at all, i.e., the same as that
described in the first paragraph of this section.  (This may be expressed
colloquially by saying that if one observes that the fastest horse is in
front then one gains no information about the rest.)  The property has in
fact been established in not only the two species but also the $n$-species
TASEP (see \cite{fm}, Proposition~6.2), using a representation of the
stationary measure based on queuing theory; a direct proof for the two
species model may be given using the two properties of second class
particles noted in the previous paragraph.  We remark that the property of
factorization around a second class particle does not extend in a direct
way to the $n$-species model \cite{fm}.

 \medskip\noindent
 2.  The fact that the measures $\nu^{\rho_0,\rho_1}$ are not product
measures gives extra structure to the local states $\mu_{x,c}$ discussed in
Section~\ref{sec:localbulk}, which are superpositions of such measures.
We note here however that as in the case of
the one component TASEP, when such a superposition occurs only on the shock
line $\alpha=\beta<1/2$, the translation invariant measures which enter
into the superposition (and which correspond to the measures on one side or
another of a shock) all have the same current.
This can be interpreted as saying that the properly averaged local current
does not fluctuate. These averages can be obtained either as long time
averages of the stochastic flux across a single bond, or as spatial
averages over an interval of length $K$, with $K \to \infty$ {\em after}
$L \to \infty$. We believe that the convergence of the average total flux
across an open system to a deterministic value, as $L\to\infty$, is a
general property of the NESS of systems like those discussed here, but do
not know how to prove this directly at the present time.  It seems
reasonable to expect similar behavior in higher dimensions and different
settings, e.g., for driven diffusive systems on a torus \cite{kls}.

 \medskip\noindent
 3.  It follows from the ``separating'' property of conditioning on the
presence of a second class particle at a given site that the distribution
under $\nu^{\rho_0,\rho_1}$ of the second class particles alone is given by
a renewal process \cite{djls,ffk}.  When the current $J_2$
vanishes, i.e., when $\rho_0=\rho_1 = (1-\rho_2)/2$, then (as noted in
Section~\ref{sec:equil}) the distribution of the distance between nearest
neighbor particles in this process has a simple exponential dependence on
$\rho_2$ which can be obtained from  a pressure ensemble, with
$p = -\log(1-\rho_2^2)$, as in region I of the open system.

Combining this expression for the pressure as a function of the density
with standard thermodynamic relations we can obtain expressions for the
chemical potential $\lambda$ and Helmholtz free energy $a$ in the uniform
infinite system of second class particles with density $\rho_2$:
\be \label{helm}
\begin{split}
\lambda(\rho_2) & = - \log \left( \frac{1-\rho_2}{1+\rho_2} \right), \\
a(\rho_2) & = (1-\rho_2) \log(1-\rho_2) + (1+\rho_2) \log(1+\rho_2).
\end{split}
\ee
 From \eqref{helm} we can obtain the large deviation function for the
probability of finding $n_2 \cL$ particles in an interval of $\cL$ lattice
sites \cite{dls}.  The large deviation for first class particles or holes
alone is of course given by the properties of the Bernoulli measure.  Large
deviation properties of the full measure have not, so far as we know, been
determined for the two species system.

 \medskip\noindent
 4. Even knowing fully the properties of the infinite system still leaves
open the problem of how fast the correlations in the vicinity of a site
$\floor{xL}$ approach those in the local measure $\mu_{x,c}$.  This may be
of particular interest in the case when $\mu_{x,c}$ is a superposition of
extremal infinite volume measures $\nu^{\rho_0,\rho_1}$ as discussed in
Section~\ref{sec:localbulk}.  We might expect the $L$ dependence in that
case, where typical density profiles differ from average ones, to be
different from that where the two coincide.  We leave this as an open
problem.

 \medskip\noindent
  5. We now briefly describe two related model systems, containing both first
and second class particles on a ring, which are intermediate between those
studied in \cite{djls} and in this paper.

 \smallskip\noindent 
5.1 Recall the ``truck'' or ``defect particle'' model \cite{derr,mal,de,be}, a
standard two species TASEP system: a single defect particle together with
(first class) particles and holes, on a ring of $L+1$ sites, can exchange
with a hole ahead of it (clockwise) at rate $\alpha$ and with a particle
behind it at rate $\beta$.  Let us add to the ring also $n$ (standard)
second class particles, which make the same exchanges as does the defect
particle but at rate 1 in each case, and which do not exchange at all with
the defect particle.  To be definite let us say that there are $n_1$ first
class particles and $n_0$ holes on the ring, with $n+n_1+n_0=L$.  Then the
stationary measure for this system is almost the same as that of our open
system: using the matrices $X_0$, $X_1$, and $X_2$ of
Section~\ref{sec:matrix}, and introducing also
$X_\delta= |V_\beta \rangle\langle W_\alpha|$, we find that a
configuration $\delta,\tau_1,\ldots,\tau_L$, where $\delta$ represents the
defect particle, has weight:
\be \label{probtruck}
  \Tr(X_\delta X_{\tau_1} \cdots X_{\tau_L})
  =  \langle W_\alpha| X_{\tau_1} \cdots X_{\tau_L} |V_\beta \rangle,
\ee
 (compare \eqref{probtau}).  The difference, of course, is that this is a
canonical ensemble and the partition function must be obtained by summing
the weights over only those configurations with the proper numbers of all
species.  This relation between this truck model on a ring of $L+1$ sites
and the two species open system studied in this paper is completely
parallel to that between the usual defect particle model and the open one
species TASEP.  We expect that details of the stationary state could be
worked out in parallel to that of the usual defect particle model, but we
have not done so.

 \smallskip\noindent
 5.2 In the second model, the ring has $N$ sites labeled by
$i \in [-N/2+1,N/2]$ and contains $N_1 = \bar{\rho}_1 N$ first class
particles, $N_2 = \bar{\rho}_2 N$ second class particles, and
$N_0 = N-N_1-N_2= \bar{\rho}_0 N$ holes. The particles jump clockwise
according to the TASEP rules given in section \ref{sec:model}, {\em except}
at one specified semi-permeable bond, say between site 0 and site 1, which
prohibits the passage of second class particles. (We can think of this
bond as a restriction in a channel).

\begin{figure}[ht!]
\includegraphics[width=6.2cm]{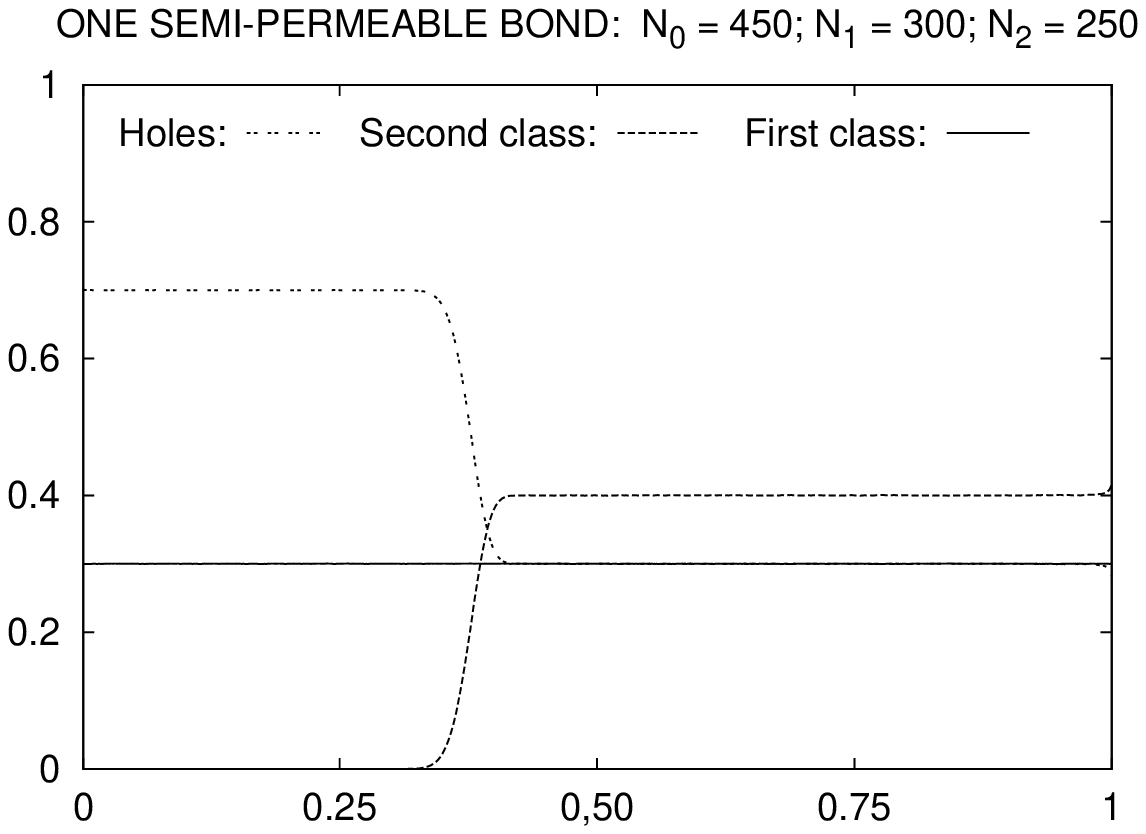}
\includegraphics[width=6.2cm]{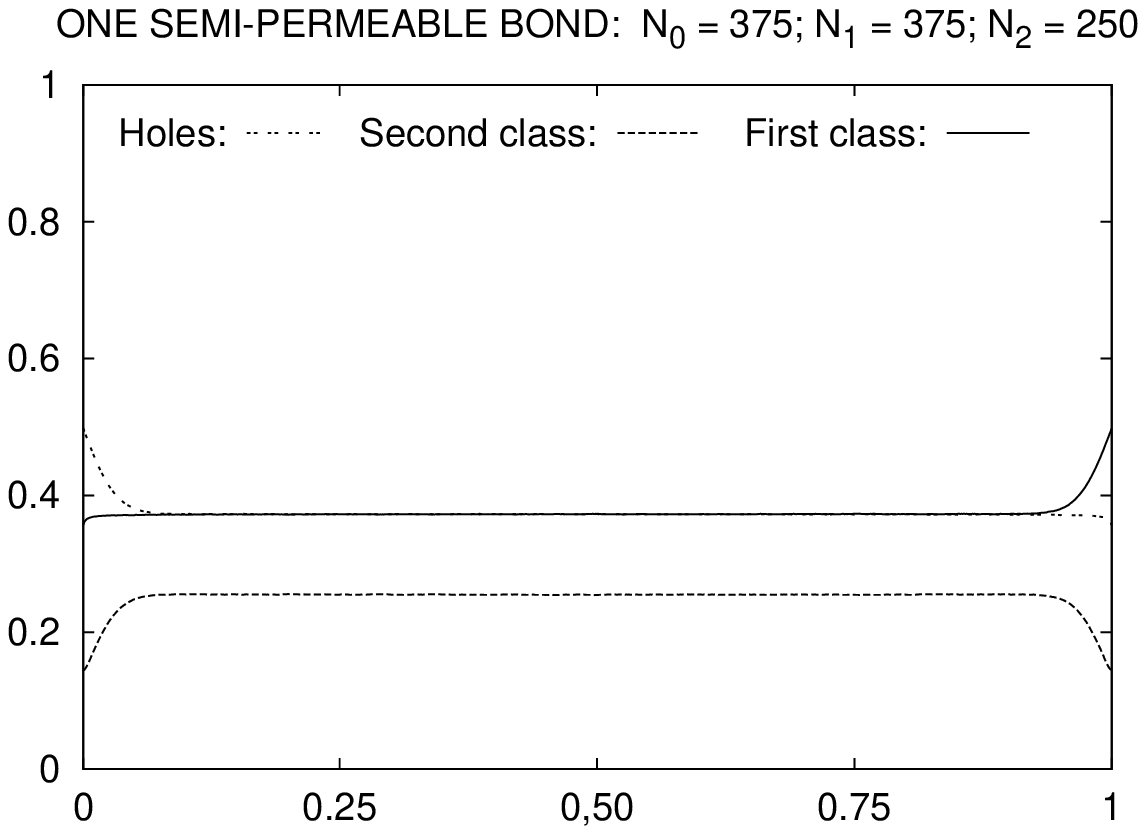}

\hbox to\hsize{\hfil(a)\hfil\hfil(b)\hfil}

\caption{Density profiles in a system with one semi-permeable bond: $L=1000$.} 
\label{fig:semiprofs}
\end{figure}

Unfortunately we do not have an exact solution for this system.  To see
what happens, however, we note that, as in the open system, the current
$J_2$ of second class particles will vanish in the stationary state.  On
the other hand, since we would have $J_2 = \rhobar_2 (\rhobar_0-\rhobar_1)$
if the system were uniform, a uniform state is possible only if
$\rhobar_1=\rhobar_0$.  If $\rhobar_1 < \rhobar_0$ then $J_2$ would be
positive in the uniform system and second class particles would drift to
the right; the upshot is that there will be a fat shock of width
$w=\bar{\rho}_2N/(1-2\bar{\rho}_1)$ containing all second class
particles at density $\rho_2 = 1-2\rho_1$ pinned to the back of the
barrier.  If $\rhobar_1>\rhobar_2$ then the fat shock of width
$w= \bar{\rho}_2N/(1-2\bar{\rho}_0)$ will be pinned to the front of
the barrier.  In the case $\rhobar_0 = \rhobar_1 = (1-\rhobar_2)/2$ the
system will be uniform.  See Figure~\ref{fig:semiprofs} for some typical
profiles in this system; note that $N_0$, $N_1$, and $N_2$ have been chosen
so that the bulk densities in Figures~(a) and (b) here are the same as
those in Figures~(a) and (e) of Figure~\ref{fig:profs}, but that the
boundary effects and finite density shock transition are noticeably
different in the two models.

Letting $N \to \infty$ with $\bar{\rho}_1,\bar{\rho}_0$ fixed would yield
an infinite system with a barrier at the bond $(0,1)$. Consider first the
case $\bar{\rho}_1<\bar{\rho}_0$. To the right of the origin there would be
no second class particles and a uniform density of first class particles
described by a product measure.  Far to the left of the barrier the state
would be $\nu^{\rhobar_0,\rhobar_0}$, i.e., a uniform state with
$\rho_1=\rho_0 =\bar{\rho}_1$ and $\rho_2 = 1-2\bar{\rho}_1$. We do not
know, however, the structure of the system just to the left of the barrier.
Similar conclusions hold for $\rhobar_1>\rhobar_0$.

\bigskip\noindent
{\bf Acknowledgments.} We thank B.~Derrida, S.~Goldstein, P.~Ferrari, and
D.~Zeilberger for valuable comments.  The work of J.L.L.~and A.A~was
supported by NSF Grant DMR-0442066 and AFOSR Grant AF-FA9550-04.  Any
opinions, findings, conclusions, or recommendations expressed in this
material are those of the authors and do not necessarily reflect the views
of the National Science Foundation.

\appendix
\section{A particular representation}\label{app:rep}

A representation of the algebra  \eqref{dereq}--\eqref{deric} which
satisfies \eqref{conseq}  may be obtained from \cite{dehp} and
\cite{djls}:
\be
 X_1 = \left( \begin{array}{cccccc}
		  1&1&0&0&. &.\\
		  0&1&1&0&& \\
		  0&0&1&1&&\\
		  0&0&0&1&.  &\\
		  . &&&&. &. \\
		  . &&&&&.
		   \end{array}
		   \hspace{0.2in} \right)
\;,\qquad
X_0 = \left( \begin{array}{cccccc}
		  1&0&0&0&. &.\\
		  1&1&0&0&& \\
		  0 &1&1&0&&\\
		  0 &0&1&1&&\\
		  . &&&.&. & \\
		  .&&&&.&.
		   \end{array}
		   \hspace{0.2in} \right)\;.
\label{DEdef}
\ee
\be
  X_2 = X_1X_0-X_0X_1 = [X_1,X_0]= \left( \begin{array}{cccccc}
		  1&0&0&0&. &.\\
		  0&0&0&0&& \\
		  0 &0&0&0&&\\
		  0 &0&0&0&&\\
		  . &&&&. & \\
		  .&&&&&.
		   \end{array}
		   \hspace{0.2in} \right)\;,
\label{DEA}
\ee
\be
\langle W_\alpha|=\left(1,\left(\frac{1-\alpha}\alpha\right),
  \left(\frac{1-\alpha}\alpha\right)^2,\ldots \right),\qquad
|V_\beta\rangle=\left(\begin{array}{c}
  1\\ \ds\frac{1-\beta}\beta \\
  \ds\left(\frac{1-\beta}\beta\right)^2\\ \vdots
		   \end{array} \right).\label{VWdef}
\ee
The exponential growth of the components of $\langle W_\alpha|$ and
$|V_\beta \rangle$ for certain values of $\alpha$ and $\beta$ in fact
causes no concern here: because we always have $n>0$, the matrix product
needed to calculate the probability of any configuration $\tau$ (see
\eqref{probtau}) will contain at least one factor $X_2$,   and using
\eqref{conseq} one can see that this implies that the corresponding matrix
element is finite.

\section{Asymptotics of the partition function}\label{app:part}

 We summarize here the
asymptotics of the partition function which are needed in
Section~\ref{sec:localbulk}.  For the case with no second class particles
\cite{dehp} we need $Z^{\alpha,\beta}$ only when $\alpha=1$ and/or
$\beta=1$:
 \be\label{z1alpha}
 Z^{\alpha,1}(j,0) = Z^{1,\alpha}(j,0) \sim  \begin{cases}
   \ds\frac{1-2\alpha}{(1-\alpha)^2}\left(\frac1{\alpha(1-\alpha)}\right)^{j},
            &\text{if $\alpha<1/2$,}\\\noalign{\smallskip}
   \ds\frac{2}{\sqrt\pi}\frac{4^j}{j^{1/2}},&\text{if $\alpha=1/2$,}\\
  \noalign{\smallskip}
   \ds\frac{\alpha^2}{\sqrt\pi(2\alpha-1)^2}\frac{4^{j+1}}{j^{3/2}},
  &\text{if $\alpha>1/2$.}\end{cases}
 \ee
 The generating  function is \cite{be}
 \be\label{genfct}
 \sum_{L=1}^\infty \lambda^LZ^{\alpha,\beta}_{L,0} 
  = \left(\frac{2\alpha}{2\alpha-1+\sqrt{1-4\lambda}}\right)
 \left(\frac{2\beta}{2\beta-1+\sqrt{1-4\lambda}}\right).
 \ee

For the model with second class particles \cite{arita}:
 \begin{itemize}

\item In region I, ($\alpha_c < \alpha,\beta$)
\be \label{denomasymp1}
Z^{\alpha,\beta}(L,n) = 
\frac{n\alpha\beta\sqrt{L^2-n^2}}{\sqrt{\pi L}((2\alpha-1)L+n)((2\beta-1)L+n)}
\left(\frac{4L^2}{L^2-n^2} \right)^{L+1}\!\! \left( \frac{L-n}{L+n} 
\right)^{n};
\ee

\item In region II, ($\alpha<\alpha_c,\alpha<\beta $)
\be
Z^{\alpha,\beta}(L,n) = \frac{\beta(1-2\alpha)}{\alpha(\beta-\alpha)} 
\left(\frac{1}{\alpha(1-\alpha)}\right)^{L+1} 
\left(\frac{\alpha}{1-\alpha}\right)^{n}\,;
\ee

\item On the boundary of regions I and II, ($\alpha_c = \alpha < \beta$)
\be
Z^{\alpha,\beta}(L,n) 
  =  \frac{\beta n(L-n)}{2L((2\beta-1)L+n)} 
  \left(\frac{4L^2}{L^2-n^2}\right)^{L+1}\left(\frac{L-n}{L+n}\right)^{n}
\ee

\item On the boundary of regions II and III, ($ \alpha=\beta<\alpha_c $)
\be
Z^{\alpha,\beta}(L,n) 
  =  \frac{(1-2\alpha)((1-2\alpha)L-n)}{(1-\alpha)^2} 
\left(\frac{1}{\alpha(1-\alpha)}\right)^{L}
 \frac{\alpha1}{(1-\alpha)}
\ee

\item At the triple point, ($\alpha_c = \alpha = \beta$)
\be \label{denomasympn}
Z^{\alpha,\beta}(L,n) 
  = \frac{n(L-n)}{2L(L+n)} \sqrt{\frac{L^2-n^2}{L\pi}} 
\left(\frac{4L^2}{L^2-n^2} \right)^{L+1} 
 \left(\frac{L-n}{L+n} \right)^n\,.
\ee

\end{itemize}

\noindent Asymptotics in Region III and on the I/III boundary are obtained
from those of Region II and the I/II boundary by exchange of $\alpha$ and
$\beta$.

\section{Finite volume corrections to density profiles}\label{sec:corrections}

We consider here again the problem of finding asymptotic values of the
density profiles, beginning with a discussion of the method of
\cite{arita}.  The partition function can be expressed as
\be \label{denomr}
Z^{\alpha,\beta}(L,n) = \frac{\alpha\beta}{\alpha-\beta} 
\left[ R(L,n,\beta) -R(L,n,\alpha) \right],
\ee
where
\be \label{rdef}
R(L,n,\alpha) =  \sum_{k=0}^{L-n} C^{L+n-1}_{L-n-k} \frac{1}{\alpha^{k+1}},\\
\ee
 with $C^{m}_{n}$ the Catalan triangle numbers \eqref{eqcattri}.  An
asymptotic analysis of \eqref{rdef} then leads, through \eqref{denomr} and
the formulas \eqref{F}--\eqref{G} for the densities, to the density
asymptotics.  In \cite{arita} the asymptotic density at position $x$ was
calculated as
 \be\label{limitbis}
  \lim_{L\to\infty}
  \E{\mu_{L,\floor{\gamma L}}^{\alpha,\beta}}{\eta_a(i_L)},
 \ee
  with $i_L=\floor{x L}$.  As observed in Section~\ref{sec:localbulk},
however, if $x$ is the location of the fixed shock in regions II or III,
and one considers limits as in \eqref{limitbis} with
$i_L=\floor{xL}+c\sqrt L$, then the limiting density value depends on $c$.
This $c$ dependence may be calculated by the methods of
Section~\ref{sec:localbulk} (see for example \eqref{rhoshock}); here we
sketch briefly an alternate and more direct method which extends the work
of \cite{arita}.

The key step is the computation of the asymptotics of $R(L,n,\alpha)$; it
is convenient to introduce $\alpha_c=(L-n)/(2L)$ (see \eqref{alphac}).  We
must determine which terms in \eqref{rdef} dominate the sum.  If we let
$L\to\infty$ at fixed $n$ and $\alpha$ there are three regimes:
(i)~$\alpha>\alpha_c$, for which the maximum of the summand is attained
when $k$ is of order $L$ and the sum can be approximated by a Gaussian
integral; (ii)~$\alpha<\alpha_c$, in which the maximum is attained when $k$
is of order $-L$ and the sum can be approximated by a geometric series; and
(iii)~$\alpha=\alpha_c$, for which the maximum occurs when $k$ is of order
$1$ and the sum can be approximated by half of a Gaussian integral.
However, there are intermediate regimes in which the sum is dominated by
terms in which $k$ is of order $\pm\sqrt L$, and it is these which generate
the finite volume density corrections that we seek.

One needs an asymptotic estimate of $R(L,n,\alpha)$ which holds for all
large $L$ and $n$.  Such an estimate is 
$R(L,n,\alpha)\sim \tilde R(L,n,\alpha)$, where
\be \label{rasymp}
\tilde R(L,n,\alpha) = \left\{ \begin{array}{ll}
\displaystyle (1-2\alpha) \left(\frac{1}{\alpha(1-\alpha)} \right)^{L+1} 
   \left( \frac{\alpha}{1-\alpha} \right)^{n}
   \Phi\left(\frac{L(1-2\alpha)-n}{\sqrt{\alpha(L+n)}}\right),\\
  \\
   & \hskip-70pt \alpha\le\alpha_c, \\
\\
\displaystyle \frac{2nL}{(n-L(1-2\alpha))\sqrt{\pi L (L^2-n^2)}}
 \left(\frac{4L^2}{L^2-n^2} \right)^{L}\\
  \\ 
\hskip20pt\displaystyle\times\left( \frac{L-n}{L+n} \right)^n 
\Psi\left(\frac{L(1-2\alpha)-n}{\sqrt{\alpha(L+n)}}\right),
 & \hskip-70pt \alpha>\alpha_c;
\end{array} \right.
\ee
 here $\Phi$ is as 1 in \eqref{erf} and
$\Psi(t)=\sqrt{2\pi}e^{t^2/2}|t|\Phi(t)$.  The asymptotic estimate
$R\sim\tilde R$ holds in the sense that for $\alpha$ and the ratio $n/L$
uniformly bounded away from $0$ and $1$ the quantity $|R/\tilde R-1|$ is
small when $L$ is large---more precisely, for any $\epsilon>0$ there is a
constant $C_\epsilon$ such that
$|R/\tilde R-1|\le C_\epsilon L^{-1/2-\epsilon}$.  We remark that the two
forms in \eqref{rasymp} in fact agree for
$\alpha_c<\alpha<\alpha_c+O(1/\sqrt{L})$.

From \eqref{denomr} and \eqref{rasymp} one obtains similarly improved
asymptotics for the partition function $Z^{\alpha,\beta}(L,n)$, and the
full density asymptotics then follows from the exact formulas of
\cite{arita} or Theorem~\ref{thm:dens}.

\end{document}